# The Flux Qubit Revisited to Enhance Coherence and Reproducibility

Fei Yan[1], Simon Gustavsson[1], Archana Kamal[1], Jeffrey Birenbaum[2,*], Adam P. Sears[3], David Hover[3], Ted J. Gudmundsen[3], Danna Rosenberg[3], Gabriel Samach[3], Steven Weber[3], Jonilyn L. Yoder[3], Terry P. Orlando[1,4], John Clarke[2], Andrew J. Kerman[3], and William D. Oliver[1,3,5]

[1]Research Laboratory for Electronics, Massachusetts Institute of Technology, Cambridge, MA 02139; [2]Department of Physics, University of California, Berkeley, CA 94720-7300; [3]MIT Lincoln Laboratory, Quantum Information and Integrated Nanosystems Group, 244 Wood Street, Lexington, MA 02420; [4]Department of Electrical Engineering and Computer Science, Massachusetts Institute of Technology, Cambridge, MA 02139; [5]Department of Physics, Massachusetts Institute of Technology, Cambridge, MA 02139

* Current address: jeffrey.birenbaum@ll.mit.edu.

**The scalable application of quantum information science will stand on reproducible and controllable high-coherence quantum bits (qubits). Here, we revisit the design and fabrication of the superconducting flux qubit, achieving a planar device with broad frequency tunability, strong anharmonicity, high reproducibility, and relaxation times in excess of 40 μs at its flux-insensitive point. Qubit relaxation times $T_1$ across 22 qubits are consistently matched with a single model involving resonator loss, ohmic charge noise, and *1/f* flux noise, a noise source previously considered primarily in the context of dephasing. We furthermore demonstrate that qubit dephasing at the flux-insensitive point is dominated by residual thermal photons in the readout resonator. The resulting photon shot noise is mitigated using a dynamical decoupling protocol, resulting in**

**$T_2 \approx 85~\mu\text{s}$, approximately the $2T_1$ limit. In addition to realizing an improved flux qubit, our results uniquely identify photon shot noise as limiting $T_2$ in contemporary qubits based on transverse qubit-resonator interaction.**

Over the past 15 years, superconducting qubits have achieved a remarkable five-order-of-magnitude increase in their fundamental coherence metrics, including the energy decay time $T_1$, the Ramsey free-induction decay time $T_2^*$, and the refocused Hahn-echo decay time $T_{2E}$. This spectacular trajectory is traceable to two general strategies that improve performance: (1) reducing the level of noise in the qubit environment through materials and fabrication improvements, and (2) reducing the qubit sensitivity to that noise through design advancements[1].

The charge qubit evolution is a quintessential example[2]. Early demonstrations (Cooper-pair box) exhibited nanosecond-scale coherence times[3]. Since then, operation at noise-insensitive bias points (quantronium)[4], the introduction of capacitive shunting (transmon)[5], the use of 2D[6] and 3D[7] resonators to modify the qubit electromagnetic environment, the development of high-Q capacitor materials and fabrication techniques[8,9], and the introduction of alternative capacitor geometries (Xmon)[10] have incrementally and collectively raised coherence times to the 10-100 μs range[10,11] and beyond[12,13]. In addition, the capacitive shunt has generally improved device-to-device reproducibility. The trade-off, however, is a significant reduction in the charge qubit intrinsic anharmonicity (*i.e.*, the difference in transition frequencies $f_{01}$ and $f_{12}$ between qubit states 0, 1 and 1, 2) to 200-300 MHz for contemporary transmons, complicating high-fidelity control and exacerbating frequency crowding in multi-qubit systems[14].

In contrast, the performance of the persistent-current flux qubit[15,16] has progressed more slowly over the past decade. Device asymmetry was identified early on to limit flux qubit coherence[17] and, since 2005, symmetric designs have generally achieved 0.5-5 μs[18,19] with a singular report of $T_{2E} = 23~\mu\text{s} \approx 2T_1$ [20]. Despite respectable performance for individual flux qubits, however, device-to-device reproducibility has remained poor. An early attempt at capacitive shunting[21] improved reproducibility, but coherence remained limited to $1-6~\mu s$ [22,23]. Recently, flux qubits embedded in 3D[24] and coplanar[25] resonators exhibited more reproducible and generally improved relaxation and coherence times: $T_1 = 6-20~\mu s$, $T_2^* = 2-8~\mu s$. Nonetheless, further improvements in these times and in reproducibility are necessary if the flux qubit is to be a competitive option for quantum information applications.

In this context we revisit the design and fabrication of the flux qubit. Our implementation, a capacitively shunted (C-shunt) flux qubit[21] coupled capacitively to a planar transmission-line resonator, exhibits significantly enhanced coherence and reproducibility while retaining an anharmonicity varying from 500 – 910 MHz in the four devices with the highest relaxation times. We present a systematic study of 22 qubits of widely varying design parameters – shunt capacitances $C_{sh} = 9-51$ fF and circulating currents $I_p = 44 - 275$ nA – with lifetimes at the flux-insensitive bias point ranging from $T_1 < 1 \mu s$ (small $C_{sh}$, large $I_p$) to $T_1 = 55 \mu s$ (large $C_{sh}$, small $I_p$). Over this entire range, the measured $T_1$ values are consistent with a single model comprising ohmic charge noise, $1/f$ flux noise, and Purcell-enhanced emission into the readout resonator. We furthermore investigated and identified quasiparticles as a likely source of observed $T_1$ temporal variation. For the highest coherence devices, the Hahn-echo decay time $T_{2E} = 40 \mu s < 2T_1$ does not reach the $2T_1$ limit, as is also often observed with transmons coupled transversally to resonators[7,10,26]. We demonstrate that this is due to dephasing caused by the shot noise of residual photons in the resonator (mean photon number $\bar{n}_0 = 0.006$), observing a lorentzian noise spectrum with a cut-off frequency consistent with the resonator decay rate. We then use Carr-Purcell-Meiboom-Gill (CPMG) dynamical decoupling to recover $T_{2CPMG} \approx 2T_1$ in a manner consistent with the measured noise spectrum.

**Results**

**C-shunt flux qubit.** Our circuits each contain two C-shunt flux qubits—with different frequencies—placed at opposite ends of a half-wavelength superconducting coplanar waveguide (CPW) resonator (Fig. 1a). The resonator, ground plane, and capacitors (Fig. 1b) were patterned from MBE-grown aluminum deposited on an annealed sapphire substrate[8] (Supplementary Note 1). We used both square capacitors (Fig. 1b) and interdigital capacitors (IDCs, not shown) coupled capacitively to the center trace of the CPW resonator to enable qubit control and readout. In a second fabrication step, the qubit loop and its three Josephson junctions (Fig. 1c) were deposited using double-angle, electron-beam, shadow evaporation of aluminum. One junction is smaller in area (critical current) by a factor $\alpha$, and each of its leads contacts one electrode of the shunt capacitor. An equivalent circuit is illustrated in Fig. 1d (Supplementary Note 2).

Varying the qubit design enables us to explore a range of qubit susceptibilities to flux and charge noise with impact on both $T_1$ and $T_2$ [21]. Compared with the conventional persistent-current flux qubit[15,16], our best C-shunt flux qubits have two key design enhancements. First, a smaller circulating current – achieved by reducing the area and critical current density of the Josephson junctions (Fig. 1c) – reduces the qubit sensitivity to flux noise, a dominant source of decoherence in flux qubits. Second, a larger effective junction capacitance – achieved by capacitively shunting the small junction (Fig. 1b) – reduces the qubit sensitivity to charge noise, and improves device reproducibility by reducing the impact of both junction fabrication variation and unwanted stray capacitance. Furthermore, the use of high-quality fabrication techniques and physically large shunt capacitors reduces the density and electric participation of defects at the various metal and substrate interfaces[1].

The system is operated in the dispersive regime of circuit quantum electrodynamics (cQED) and is described by the approximate Hamiltonian[27]

$$\mathcal{H}_{\text{disp}} \approx \hbar \omega_q (\Phi_b) \hat{\sigma}_z / 2 + \hbar \omega_r (\hat{n} + 1/2) + 2\hbar \chi (\Phi_b)(\hat{n} + 1/2) \hat{\sigma}_z / 2, \qquad (1)$$

where the three terms are respectively the qubit (represented as a two-level system), resonator, and qubit-resonator interaction Hamiltonians, $\hat{\sigma}_z$ is the Pauli operator defined by the qubit energy eigenbasis, $\omega_r$ is the resonator angular frequency and $\hat{n}$ is the resonator photon-number operator. The qubit angular frequency $\omega_q(\Phi_b)$ is set by the magnetic flux bias $\Phi_b$, measured relative to an applied flux $(m+1/2)\Phi_0$ where $m$ is an integer and $\Phi_0$ is the superconducting flux quantum, and attains its minimum value $\omega_q(0) \equiv \Delta$ at the flux-insensitive point $\Phi_b = 0$. The quantity $\chi(\Phi_b)$ is the qubit-state-dependent dispersive shift of the resonator frequency, which is used for qubit readout. In the Supplementary Notes 3-5, we discuss further the two-level-system approximation for the C-shunt flux qubit, an approximate analytic treatment which goes beyond Eq.(1), and the numerical simulation of the full qubit-resonator Hamiltonian used to make quantitative comparisons to our data.

$T_1$ **relaxation and noise modeling.** We begin by presenting the $T_1$ characterization protocol for the device in Fig. 1. We first identify the resonator transmission spectrum (Fig. 2a, top panel) by scanning the readout-pulse frequency $\omega_{ro}$ about the bare resonator frequency $\omega_r/2\pi \approx 8.27$ GHz. Using standard cQED readout, qubit-state discrimination is achieved by monitoring the qubit-state-dependent transmission through the resonator[27]. Next, we add a qubit driving pulse of sufficient duration to saturate the ground-to-excited-state transition and sweep the pulse frequency $\omega_d$ (Fig. 2a, bottom panel). The resulting spectra for qubits A and B (Fig. 1a) exhibit minima $\Delta_A/2\pi \approx 4.4$ GHz and $\Delta_B/2\pi \approx 4.7$ GHz at the qubit flux-insensitive points and increase with magnetic flux (bias current) away from these points. Finally, using a single $\pi$-pulse to invert the qubit population, we measure the $T_1$ relaxation of qubit A ($T_1 = 44\mu s$) and qubit B ($T_1 = 55\mu s$) at their flux-insensitive points (Fig. 2b). High-power spectroscopy (see Supplementary Note 6) reveals transitions amongst the first four qubit energy levels that are well matched by simulation, and identifies anharmonicities of 500 MHz in the two measured devices.

Using this protocol, we investigated 22 C-shunt flux qubits from five wafers (fabrication runs), spanning a range of capacitance values ($C_{sh} = 9 - 51$ fF) and qubit persistent currents ($I_p = 44 - 275$ nA) and featuring two capacitor geometries (interdigital and square). The junction critical currents were adjusted to maintain $\Delta/2\pi \approx 0.5 - 5$ GHz (see Supplementary Note 7).

The data were analyzed using simulations of the full system Hamiltonian and a Fermi's golden rule expression for the exited state decay rate[21],

$$\frac{1}{T_1} = \sum_\lambda 2 \frac{\left|\langle e|\hat{D}_\lambda|g\rangle\right|^2}{\hbar^2} S_\lambda(\omega_q), \qquad (2)$$

where $|g\rangle$ ($|e\rangle$) indicates the qubit ground (excited) states, and the sum is over four decay mechanisms: flux noise in the qubit loop, charge noise on the superconducting islands, Purcell-enhanced emission to the resonator mode, and inelastic quasiparticle tunneling through each of the three junctions. The operator $\hat{D}_\lambda$ is a transition dipole moment, and $S_\lambda(\omega_q)$ is the symmetrized noise power spectral density for the fluctuations which couple to it. For example, $\hat{D}_\Phi$ is a loop current operator for flux noise $S_\Phi(\omega)$, and $\hat{D}_Q$ is an island voltage operator for charge noise $S_Q(\omega)$ (Supplementary Note 8 and 9).

We considered both $S_\lambda(\omega) \propto 1/\omega^\gamma$ (inverse-frequency noise) and $S_\lambda(\omega) \propto \omega$ (ohmic noise) – the two archetypal functional forms of noise in superconducting qubits[20,28-33] – for our magnetic flux and charge noise models, and used the frequency dependence of $T_1$ for specifically designed devices to distinguish between them. While the following results are presented using symmetrized power spectral densities, we are careful to account for the distinction between classical and quantum noise processes in making this presentation (Supplementary Note 9).

For example, in Fig. 3a, Qubit C ($C_{sh} = 9$ fF) has a large persistent current ($I_p = 275$ nA) and a small qubit frequency ($\Delta_c/2\pi = 0.82$ GHz), making it highly sensitive to flux noise. Consequently, the measured $T_1$ is predominantly limited by flux noise over a wide frequency range. This $T_1$-trend constrains the flux noise model to the form $S_\Phi(\omega) \equiv A_\Phi^2 (2\pi \times 1 \text{ Hz}/\omega)^\gamma$ over the range 0.82 – 3 GHz (black dashed line, Fig. 3a). For comparison, the functional form for ohmic flux noise (grey dashed line), scaled to match $T_1$ at $\Delta_c/2\pi = 0.82$ GHz (green dot), is clearly inconsistent with all other data over this frequency range. The noise parameters $A_\Phi^2 = (1.4 \ \mu\Phi_0)^2/\text{Hz}$ and $\gamma = 0.9$ used to match the data in Fig. 3a are derived from independent measurements – Ramsey interferometry[31] and $T_{1\rho}$ noise spectroscopy[32] (Supplementary Note 10) – made at much lower frequencies in the context of classical noise related to qubit dephasing (Fig 3b). These values are commensurate with earlier work on qubits[20,31-33] and dc Superconducting QUantum Interference Devices (SQUIDs)[34]. The consistency between the magnitude and slope of the flux noise power spectra, spanning more than twelve decades in frequency – millihertz to gigahertz – is remarkable, made even more so by the fact that the data in Fig. 3b were measured with a different device (qubit B, Fig.3c).

In contrast, Qubit B ($C_{sh} = 51$ fF) has a much smaller persistent current ($I_p = 49$ nA) and larger qubit frequency ($\Delta_B/2\pi = 4.7$ GHz). Its value of $T_1$ is most strongly influenced by charge noise (magenta dashed line, Fig. 3c) in the 5.0 – 6.5 GHz range, consistent with an ohmic charge noise model of the form $S_Q(\omega) \equiv A_Q^2 \omega/(2\pi \times 1 \text{ GHz})$ with the parameter $A_Q^2 = (5.2 \times 10^{-9} e)^2/\text{Hz}$ adjusted to match the data. In addition to flux and charge noise, the predicted value of $T_1$ due to Purcell loss (light blue dashed line) is also included in Fig. 3a and 3c and involves no free parameters (see Supplementary Note 8). The resulting net value of $T_1$ due to all three mechanisms (inverse-frequency flux noise, ohmic charge noise, and Purcell loss) is indicated with a red solid line and is in relatively good agreement with the ceiling of measured $T_1$ values. As we describe below, quasiparticles are responsible for reducing the $T_1$ below this ceiling.

Using these models, Fig. 3d shows a comparison of the measured and predicted $T_1$ values for all 22 qubits. The flux noise model (from Figs. 3a and 3b) is applied to all qubits, and the Purcell loss is included with no free parameters. For the charge noise model, to achieve agreement across all devices, it was necessary to use $A_{Q,\text{SQ}}^2 = (5.2 \times 10^{-9} e)^2 / \text{Hz}$ for square capacitors (from Fig. 3b) and $A_{Q,\text{IDC}}^2 = (11.0 \times 10^{-9} e)^2 / \text{Hz}$ for interdigital capacitors (IDCs), presumably reflecting the larger electric participation of the surface and interface defects for the IDC geometry[1]. The agreement is noteworthy, given that these qubits span a wide range of designs across five fabrication runs (see Supplementary Note 7).

We note that inverse-frequency charge noise was incompatible with these data over the entire frequency range investigated (not shown), implying that the cross-over between inverse-frequency and ohmic charge noise occurred at a frequency below 0.82 GHz. However, while ohmic flux noise $S_\Phi(\omega) \propto \omega$ was inconsistent with $T_1$ over the frequency range 0.82 – 3 GHz, its functional form is plausibly consistent with data above 3 GHz when appropriately scaled (upper dashed grey line, Fig. 3a) and, therefore, cannot be conclusively distinguished from ohmic charge noise. Although the best agreement across all 22 qubits (Fig. 3d) did not require ohmic flux noise, we could not rule out its presence in the 3-7 GHz range. In Supplementary Note 11, we compare models that use ohmic charge noise (as in Fig. 3) and ohmic flux noise. Differentiating between such charge and flux noise at higher frequencies will be the subject of future work. Indeed, for both ohmic flux noise $S_\Phi(\omega) \propto \omega$ and inverse-frequency charge noise $S_Q(\omega) \propto 1/\omega$, it is certainly possible (even expected) that the former (latter) dominates the flux (charge) noise at sufficiently higher (lower) frequencies.

The measured data for qubit B (Fig. 3c) exhibit fluctuations in the range $T_1 = 20 - 60\,\mu\text{s}$ for qubit frequencies $\omega_q / 2\pi = 4.7 - 6.5$ GHz. To investigate their temporal nature, we measured $T_1$ repeatedly at the qubit flux-insensitive point $\omega_q / 2\pi = \Delta_B / 2\pi = 4.7$ GHz over a 10-hour period and collected the data into sets of 50 individual decay traces. Figures 4a and 4b show the results of two such experiments, with set 2 being taken approximately ~17 h after set 1. The average of all traces from set 1 exhibits a purely exponential decay, whereas the corresponding average for set 2 exhibits a faster short-time decay and clear non-exponential behavior (Fig. 4a). Histograms of the $T_1$ values for individual traces exhibit a tight, Gaussian-shaped distribution centered at $55\,\mu\text{s}$ for set 1 and a broader, quasi-uniform distribution centered near $45\,\mu\text{s}$ for set 2. Over the course of several weeks, we observed transitions between these two characteristic modes of behavior every few days for this device[35].

We attribute the temporal fluctuations and non-exponential decay function to excess quasiparticles – above the thermal equilibrium distribution – near the qubit junctions[36-39]. Following Ref. 40, we define $\bar{T}_{1qp}$ as the average relaxation time associated with a single quasiparticle and take the quasiparticle number $n_{qp}$ to be Poisson-distributed with mean value $\bar{n}_{qp}$. This results in a qubit polarization decay function,

$$\langle P_e(t) \rangle = e^{\bar{n}_{qp}(\exp(-t/\bar{T}_{1qp})-1)} e^{-t/T_{1R}}, \qquad (3)$$

where $T_{1R}$ captures the residual exponential decay time in the absence of quasiparticles ($\bar{n}_{qp} = 0$). The non-exponential decay function observed for set 2 is well described by Eq. (3) [black line in Fig. 4a] with fitting parameters $\bar{n}_{qp} = 0.26$, $\bar{T}_{1qp} = 23 \mu s$ and $T_{1R} = 60 \mu s$.

We use a quantum treatment of quasiparticle tunneling to model the impact of single quasiparticles on the $T_1$ of qubit B (Supplementary Note 8). Using a quasiparticle density $x_{qp} = 4 \times 10^{-7}$ (per superconducting electron), the calculated $\bar{T}_{1qp}$ recovers the fitted value $\bar{T}_{1qp} = 23 \mu s$ at the flux-insensitive point. Both $\bar{T}_{1qp}$ and $x_{qp}$ are comparable to the quasiparticle-induced relaxation rates and quasiparticle density reported for similar devices[24,41]. The shaded region in Figs. 3a and 3c indicates the range of predicted $T_1$ in the presence of $\bar{n}_{qp} = 0-1.0$ quasiparticle. Most $T_1$ data lie within this region, supporting the hypothesis that their scatter (particularly for qubit B in Fig. 3c) and the observed temporal $T_1$ variation (Fig. 4b) arise from the common mechanism of quasiparticle tunneling. Additionally, the residual relaxation time $T_{1R}$ for set 2 is similar to the exponential time constant obtained for set 1, indicating an underlying consistency in the noise models between the two data sets in the absence of quasiparticles. Unlike qubit B, qubit C consistently exhibited an exponential decay function (Fig. 4c) with little temporal variation (Fig. 3a, 4d), indicating that quasiparticles did not strongly influence this device.

The results of Fig. 3 and Fig. 4 demonstrate clearly that $1/f$-type flux noise is the dominant source of qubit relaxation for frequencies below 3 GHz. To further strengthen this claim, it is instructive to compare relaxation times for qubits with similar frequencies and shunting capacitances, but where the persistent current (and thereby the sensitivity to flux noise) differs. We find that by reducing $I_p$ from 170 nA to 60 nA, we improve the measured $T_1$ from 2.3 to 12 us (see qubits 11 and 13 in Supplementary Table 1 in Supplementary Note 7).

**Pure dephasing and thermal photon noise.** We now address the transverse relaxation time $T_2$ and our ability to refocus coherent dephasing errors. Efficient refocusing implies that $T_2$ is limited entirely by $T_1$, since $1/T_2 = 1/2T_1 + 1/T_\varphi$, where $T_\varphi$ is the dephasing time. Generally, $T_2$ is maximal at the flux-insensitive point for conventional flux qubits[18-20], and the device reported in Ref. 20 was efficiently refocused with a single echo pulse ($T_{2E} = 23\ \mu s \approx 2T_1$). In the current work, however, a single refocusing pulse is no longer completely efficient ($T_{2E} < 2T_1$). This suggests that an additional, higher-frequency noise channel has been introduced. Unlike the device in Ref. [20], which was coupled to a dc SQUID for readout, our C-shunt flux qubits are transversally coupled to a resonator (Fig. 1). Such inefficient refocusing is also reported for transmons similarly coupled to resonators[7,10,26].

As we show below, the main source of dephasing in C-shunt flux qubits biased at their flux-insensitive point is photon-number fluctuations (shot noise) in the resonator, which vary the qubit frequency via the ac Stark effect (as in the transmon case[26,27]). Given a small thermal-photon population $\bar{n} = 1/(e^{\hbar\omega_r/k_B T} - 1) \ll 1$ in the resonator (see Supplementary Note 12), the photon-induced frequency shift $\Delta_{\text{Stark}}^{\text{th}}$ and dephasing rate $\Gamma_\varphi^{\text{th}}$ of the qubit are[42]

$$\Delta_{\text{Stark}}^{\text{th}} = \eta 2\chi \bar{n}\ , \qquad (4)$$

$$\Gamma_\varphi^{\text{th}} = \eta \frac{4\chi^2}{\kappa}\bar{n}\ . \qquad (5)$$

The factor $\eta = \kappa^2/(\kappa^2 + 4\chi^2)$ effectively scales the photon population seen by the qubit due to the interplay between the qubit-induced dispersive shift of the resonator frequency $\chi$ and the resonator decay rate $\kappa$. Both the strong dispersive ($2\chi \gg \kappa$) and weak dispersive ($2\chi \ll \kappa$) regimes have been previously addressed[26,43,44]. Here, we use qubit B to focus primarily on the intermediate dispersive regime ($2\chi/2\pi = 0.9$ MHz, $\kappa/2\pi = 1.5$ MHz, see Fig. 5a) relevant for high-fidelity qubit readout[45].

We begin by intentionally injecting additional thermal photons $\bar{n}_{\text{add}}(P_{\text{add}})$ into the resonator from an external noise generator with power $P_{\text{add}}$ (Fig. 5b and Supplementary Note 2). In the small-$\bar{n}_{\text{add}}$ limit, the measured qubit spectrum exhibits a linear relationship between the effective qubit frequency $\omega_q' = \omega_q + \Delta_{\text{Lamb}} + \Delta_{\text{Stark}}^{\text{th}}(\bar{n}_{\text{add}})$ and the generator power $P_{\text{add}}$ (Figs. 5c and 5d). For completeness, we have included the Lamb shift $\Delta_{\text{Lamb}}$, a fixed frequency offset due to the resonator zero-point energy. Combining the extracted slope with Eq. (4), we calibrate the dependence of the added-photon population $\bar{n}_{\text{add}}$ (in the resonator) on the generator power $P_{\text{add}}$.

Next, we measure the Hahn-echo dephasing rate for several photon populations using the calibrated $\bar{n}_{add}(P_{add})$. All echo traces (Fig. 5e) feature exponential decay rates $\Gamma_{2E} = 1/T_{2E}$, indicating little (if any) impact from $1/f$ noise (charge, flux, …) and consistent with photon shot noise featuring a short correlation time $1/\kappa \ll T_{2E}$. The extracted pure dephasing rate $\Gamma_{\varphi E} = \Gamma_{2E} - \Gamma_1/2$ scales linearly with photon population $\bar{n}_{add}(P_{add})$ (Fig. 5f). The extracted slope agrees with Eq. (5) to within $5\%$. The non-zero dephasing rate at $\bar{n}_{add} = 0$ corresponds to a residual photon population $\bar{n}_0 = 0.006$, equivalent to an effective temperature $T_{eff} = 80$ mK. By comparison, the qubit effective temperature determined from its first excited-state population is 35 mK[13].

To confirm that the noise arises from residual thermal photons, we directly measure the noise power spectral density (PSD) using the $T_{1\rho}$ (spin-locking) method[32]. This method (inset Fig. 6a) collinearly drives the qubit along the Y-axis with a long Y pulse, which "locks" the qubit state in the rotating frame. Measuring the qubit relaxation rate in the rotating frame, $\Gamma_{1\rho}(\Omega_{Rabi}) = S_z(\Omega_{Rabi})/2 + \Gamma_1/2$, effectively samples the noise PSD $S_z(\omega)$ seen by the qubit at the locking (Rabi) frequency $\Omega_{Rabi}$ (see Supplementary Note 10). By varying the locking drive amplitude, which is proportional to $\Omega_{Rabi}$, we sample the noise spectrum over the range $\omega/2\pi = 0.1 - 100$ MHz (Fig. 6a). Below 10 MHz, the resolved noise spectra for all $\bar{n}_{add}$ (including $\bar{n}_{add} = 0$) have similar shapes: flat (white) at low frequencies with a 3-dB high-frequency cut-off at the resonator decay rate $\omega = \kappa$. This form is consistent with the expected lorenzian PSD for thermal-photons in a resonator as seen by the qubit (see Supplementary Note 10),

$$S_z(\omega) = (2\chi)^2 \frac{2\eta \bar{n} \kappa}{\omega^2 + \kappa^2} \ , \qquad (6)$$

which includes the dispersive coupling $\chi$ and the filtering factor $\eta$ [see Eqs. (4,5)]. Equation (6) agrees with the measured PSDs for all photon populations $\bar{n} = \bar{n}_{add} + \bar{n}_0$, with the residual photon number $\bar{n}_0$ extracted from Eq. (6). This agreement eliminates the driving or readout field as the source of the residual photons, because such coherent-state photons follow Poisson statistics with a resulting cut-off frequency $\kappa/2$ (one-half the observed value)[46,47].

Finally, we apply dynamical decoupling techniques to validate the functional form of the measured noise PSD and to recover $T_2 \approx 2T_1$. We use the Carr-Purcell-Meiboom-Gill (CPMG, inset Fig. 6b) pulse sequence, comprising a number $N_\pi$ of equally spaced $\pi$-pulses. The application of $\pi$-pulses in the time domain can be viewed as a bandpass filter in the frequency domain which shapes the noise spectra seen by the qubit[21,48-50]. Since the filter passband is centered at a frequency inversely related to the temporal spacing $\Delta\tau$ between adjacent pulses, increasing $N_\pi$ for a fixed sequence length will shift this passband to higher frequencies (see Supplementary Note 13).

Figure 6b shows the measured CPMG decay time $T_{2\text{CPMG}}$ vs. $\pi$-pulse number $N_\pi$ with no added noise ($\bar{n}_{\text{add}} = 0$). From $N_\pi = 1$ (Hahn echo) to $N_\pi = 100$, the decay time $T_{2\text{CPMG}}$ remains near $40\,\mu\text{s}$, consistent with the white-noise (flat) portion of the noise PSD in Fig. 6a. Above $N_\pi = 100$, the passband frequency traverses the cut-off region of the PSD and, as the integrated noise level decreases, $T_{2\text{CPMG}}$ rises. For $N_\pi > 1000$, the refocussing becomes efficient with $T_{2\text{CPMG}} \approx 85\,\mu\text{s} \approx 2T_1$. The close correspondence between the noise spectral density in Fig. 6a and the mitigation of that noise by CPMG in Fig. 6b strongly supports our methods and interpretations.

**Discussion**
The C-shunt flux qubit is a planar device with broad frequency tunability, relatively strong anharmonicity and high reproducibility, making it well suited to both gate-based quantum computing and quantum annealing. The anharmonicity can be significantly higher than that of transmon qubits, allowing for faster (even subnanosecond[51,52]) control pulses and reduced frequency crowding in multi-qubit systems. The addition of a high-quality-factor shunt capacitance to the flux qubit, together with a reduced qubit persistent current, has enabled us to achieve values of $T_1$ as high as 55 μs at the qubit flux insensitive point. We are able to account for measured $T_1$ values across 22 qubits with a single model involving ohmic charge noise, $1/f$ flux noise, and the Purcell effect, with temporal variation in $T_1$ explained by quasiparticle tunnelling. Based on this model, we anticipate further design optimization leading to even higher coherence will be possible. Finally, we used spin-locking to directly measure the photon shot noise spectral density, and we verified its functional form using a CPMG pulse sequence to reach a $T_2$ of 85 μs—limited by $2T_1$—at the flux insensitive point. These measurements identify photon shot noise as the dominant source of the observed dephasing, and have direct implications for any qubit in which the readout involves its transverse coupling to a resonator.

The role of high-frequency $1/f$ flux noise in qubit relaxation is intriguing. Our $T_1$ data and their frequency dependence across 22 different qubits strongly support the conclusion that $1/f$ flux noise contributes to qubit relaxation up to at least 3 GHz in our devices. Above 3 GHz, there is some ambiguity between ohmic flux and ohmic charge noise, and clarifying the roles of these respective noise sources is the subject of future work. A detailed understanding of such a broadband $1/f$-flux noise mechanism and its transition from classical to quantum behaviour is of great practical interest and awaits theoretical explanation.

**Acknowledgements**

We gratefully acknowledge A. Blais, A. Clerk, and Y. Nakamura for useful discussions and P. Baldo, V. Bolkhovsky, G. Fitch, P. Murphy, B. Osadchy, K. Magoon, R. Slattery, and T. Weir at MIT Lincoln Laboratory for technical assistance. This research was funded in part by the Office of the Director of National Intelligence (ODNI), Intelligence Advanced Research Projects Activity (IARPA) and by the Assistant Secretary of Defense for Research & Engineering via MIT Lincoln Laboratory under Air Force Contract No. FA8721-05-C-0002; by the U.S. Army Research Office Grant No. W911NF-14-1-0682; and by the National Science Foundation Grant No. PHY-1415514. The views and conclusions contained herein are those of the authors and should not be interpreted as necessarily representing the official policies or endorsements, either expressed or implied, of ODNI, IARPA, or the US Government.

**Author contributions**

F.Y., A.K., S.G., J.B, A.S. and D.H. performed the experiments.  S.G., A.S., A.K., D.H., A.J.K., and

W.D.O. designed; and T.G., J.Y., and J.B. fabricated the devices. A.J.K., S.G., A.K., F.Y., D.H., D.R., G.S., S.W. performed device simulations. S.G., T.P.O., J.C., A.J.K., and W.D.O. supervised the project. All authors contributed to the conception, execution and interpretation of the experiments.

## Additional information

The authors declare no competing financial interests. Supplementary information accompanies this paper on [www.nature.com/naturephysics](www.nature.com/naturephysics). Reprints and permissions information is available online at hyyp://www.nature.com/repritns. Correspondence and requests for materials should be addressed to W.D.O. ([oliver@ll.mit.edu](oliver@ll.mit.edu)).

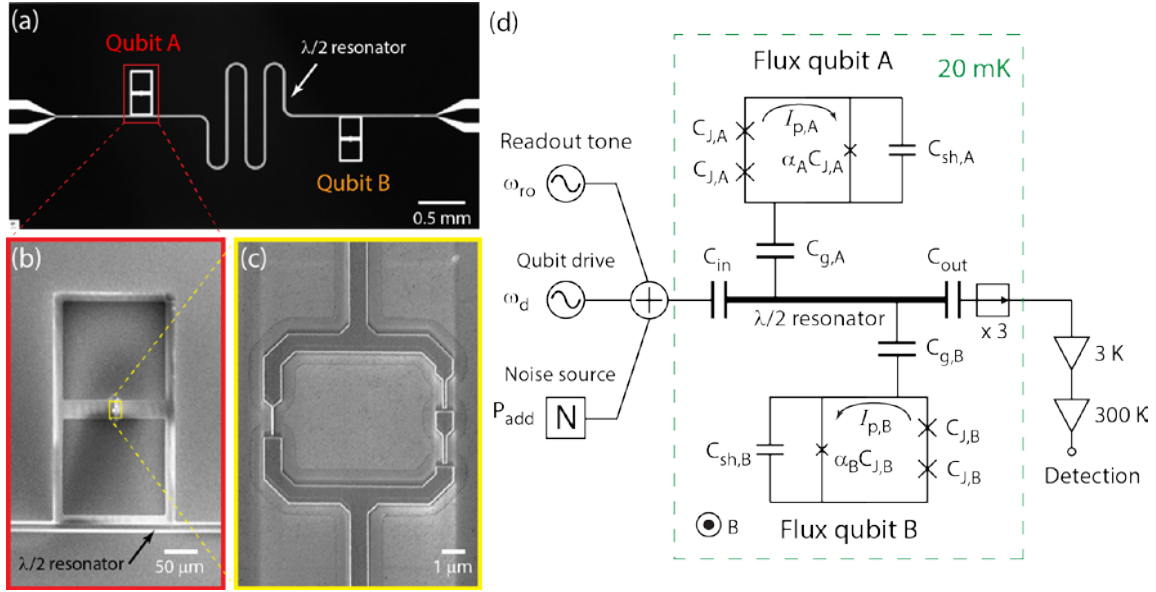

**Figure 1 | C-shunt flux qubit.**

(a) Optical micrograph of the 2.5 x 5.0 mm$^2$ chip, aluminum (black) on sapphire substrate (white, where the aluminum has been etched away), featuring two qubits (A and B) and a $\lambda/2$ coplanar waveguide resonator ($\omega_r/2\pi = 8.27$ GHz). The scale bar corresponds to 0.5 mm. (b) SEM image of the shunt capacitor ($C_{sh,A} = 51$ fF) for qubit A. Each square plate of the capacitor is 200 × 200 μm$^2$. The lower plate capacitively couples the qubit to the $\lambda/2$ resonator. The scale bar corresponds to 50 μm. (c) Magnified view of the shadow-evaporated qubit loop and its three Josephson junctions. The left junction area is smaller by a factor $\alpha_A = 0.42$. The scale bar corresponds to 1 μm. (d) Device and measurement schematic. Experiments are performed in a dilution refrigerator at 20 mK. A global magnetic field $B$ provides a magnetic flux bias $\Phi_b$ to each qubit. A qubit drive tone ($\omega_d$), readout tone ($\omega_{ro}$), and externally generated noise ($P_{add}$, see Figs. 5 and 6) enter the $\lambda/2$ resonator defined by capacitances $C_{in}$ and $C_{out}$. The resonator is capacitively coupled ($C_{g,A/B}$) to qubits A and B. The qubit junctions ("x") have internal capacitance, $C_{J,A/B}$ and $\alpha_{A/B}C_{J,A/B}$, and are externally shunted by capacitance $C_{sh,A/B}$. Each qubit loop supports a circulating persistent current $I_{p,A/B}$.

Readout signals at the resonator output pass three isolators ("$\rightarrow$"), are amplified at cryogenic and room temperatures, and subsequently detected. See supplementary online material for more information.

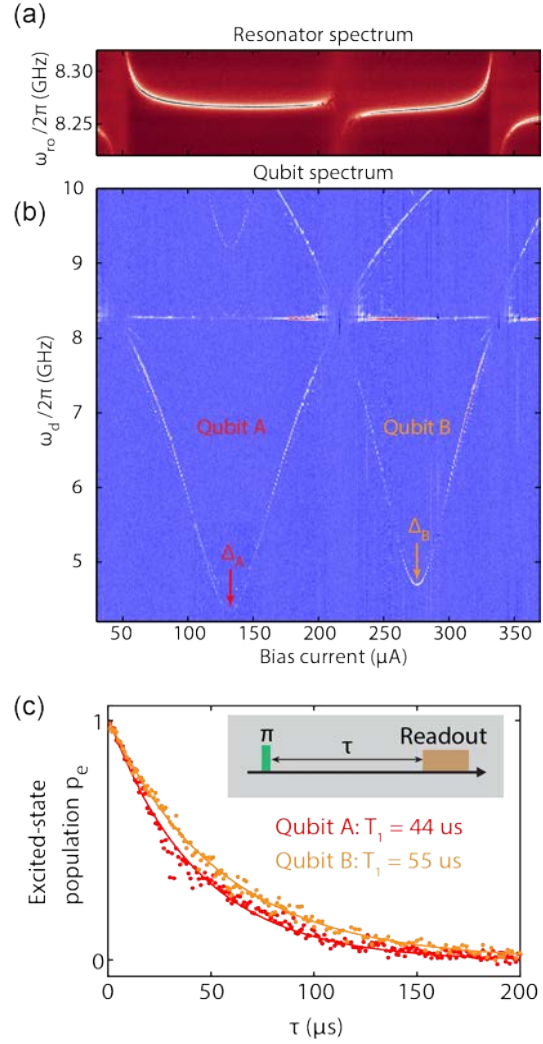

**Figure 2 | Spectroscopy and $T_1$ of two capacitively shunted flux qubits.**

**(a-b)** Resonator and qubit spectra vs. bias current used to induce the global magnetic field *B*. The qubit transition frequencies $\omega_q/2\pi$ have minima $\Delta_A/2\pi = 4.36$ GHz and $\Delta_B/2\pi = 4.70$ GHz at the qubit flux-insensitive points, which are intentionally offset in bias current (magnetic flux) by using different qubit-loop areas.

**(c)** Energy decay functions of qubits A and B measured at their respective degeneracy points using the inversion-recovery pulse sequence (inset). Solid lines are exponential fits with decay constant $T_1$.

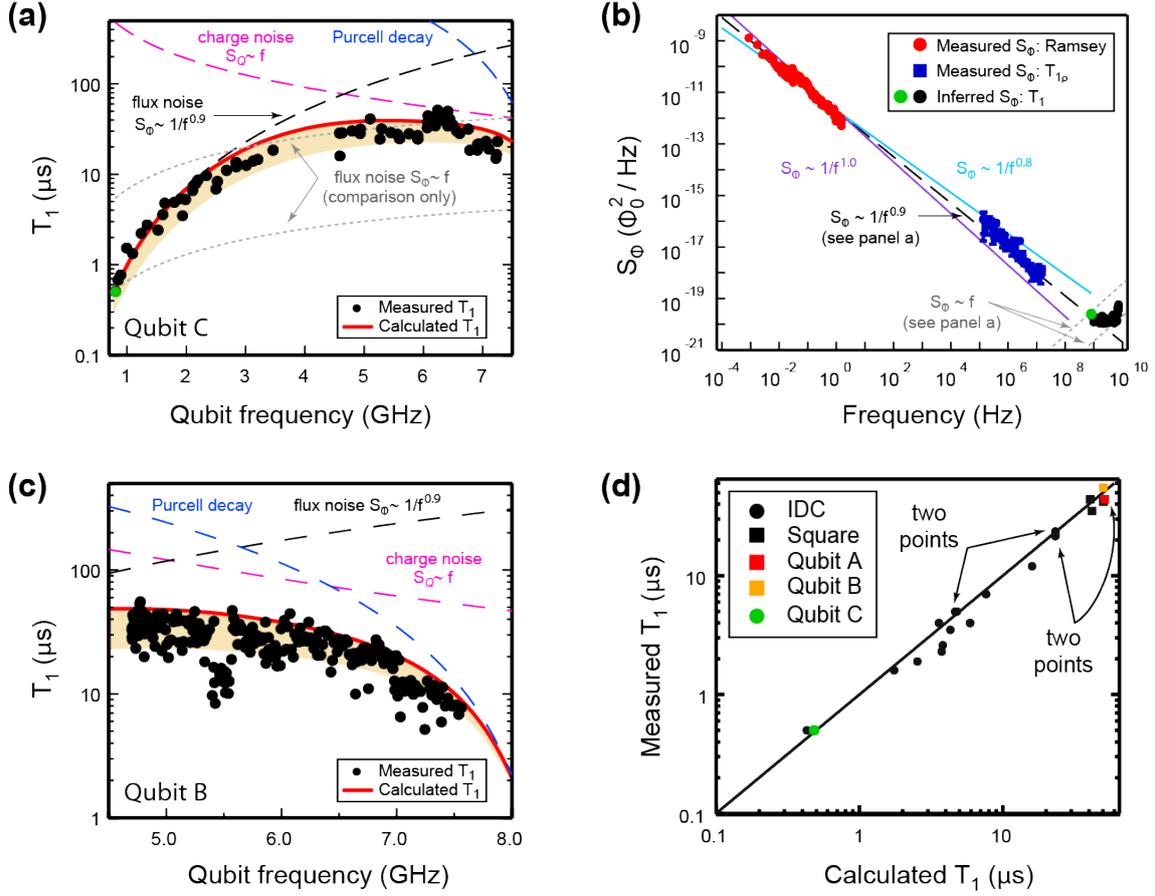

**Figure 3 | T$_1$ variation with qubit frequency and noise modeling.**

**(a)** Energy relaxation time $T_1$ vs. qubit frequency for Qubit C ($C_{\text{sh,C}} = 9$ fF, $I_{\text{p,C}} = 275$ nA, $\Delta_C/2\pi = 0.82$ GHz) plotted with simulated $T_1$ values for individual (dashed lines) and aggregate (solid line) charge, flux, and Purcell noise mechanisms. Absence of data around 4 GHz is related to an ancillary qubit level crossing the readout resonator, prohibiting qubit readout, and is not a systematic issue. Qubit C is limited by flux noise below about 4.5 GHz. For comparison, the functional form for ohmic flux noise (grey dotted line) is incompatible with the data below 3 GHz; above 3 GHz, its role cannot be readily distinguished from charge noise (see text). Shaded region indicates the range of predicted $T_1$ in the presence of 0 – 1.0 quasiparticle.

**(b)** Flux noise spectroscopy performed on Qubit B using Ramsey interferometry (red) and T$_{1\rho}$ spin locking (blue) to determine parameters $A_\Phi^2 = (1.4~\mu\Phi_0)^2/\text{Hz}$ and $\gamma = 0.9$ for the inverse frequency flux noise (black dashed line) for qubit C (panel a). Green and black dots: inferred ohmic flux noise $S_\Phi$ based on measured $T_1$ in panel a.

**(c)** Energy relaxation time $T_1$ vs. qubit frequency for Qubit B ($C_{sh,B} = 51$ fF, $I_{p,B} = 49$ nA, $\Delta_B/2\pi = 4.70$ GHz). $T_1$ is sensitive predominantly to ohmic charge noise within 5-6.5 GHz range. Scatter in $T_1$ is attributed to quasiparticle fluctuations. Cluster of lower $T_1$ values near 5.5 GHz is due to interaction with the $f_{12}$ transition. Shaded region indicates the range of predicted $T_1$ in the presence of $0 - 1.0$ quasiparticle.

**(d)** $T_1$-values for 22 qubits with widely varying design parameters, measured at their degeneracy points, and plotted against predicted $T_1$ values (dashed line) determined from numerical simulations using a single model with fixed noise levels (see main text). Practically indistinguishable data points (eight in total) are indicated with arrows.

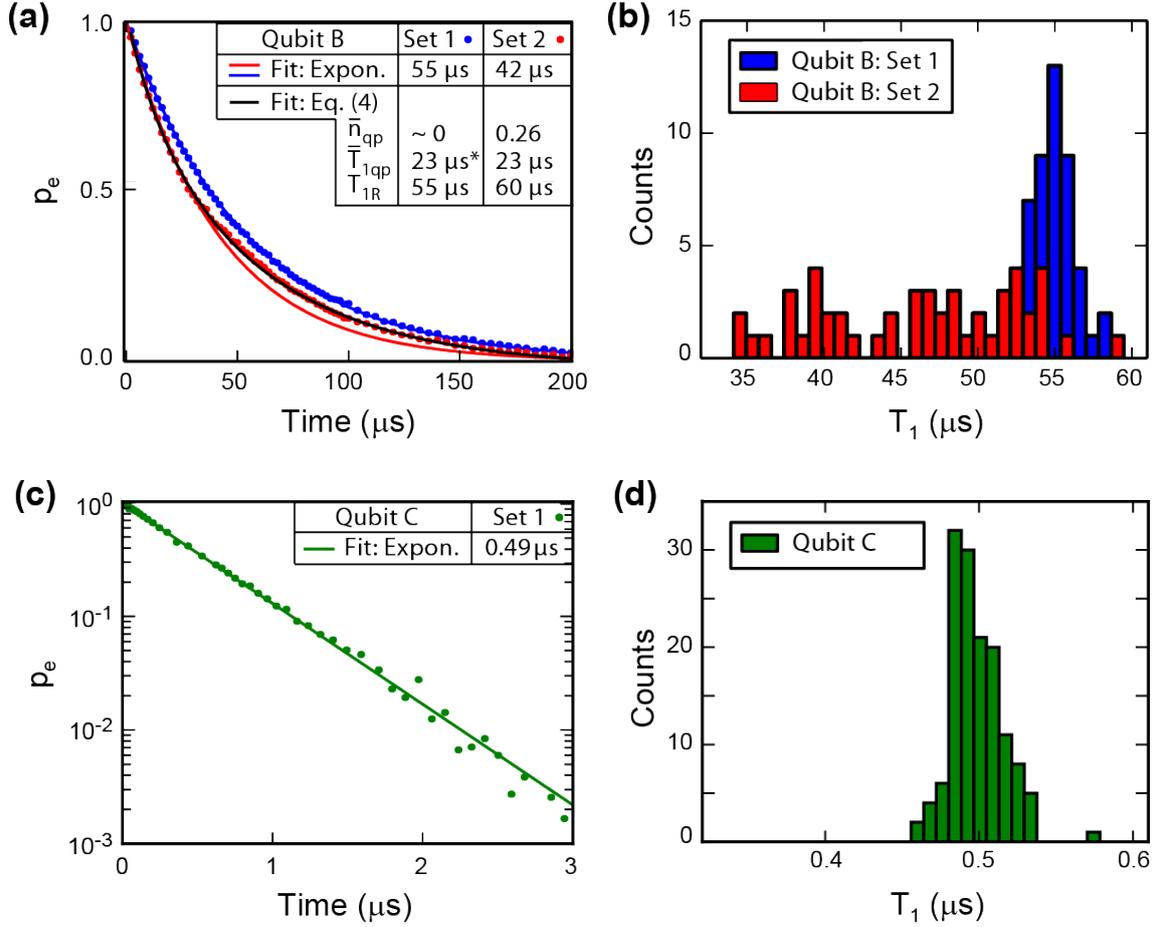

**Figure 4 | $T_1$ temporal variation and quasiparticles.**

**(a)** Energy relaxation measurements (Set 1 and Set 2) at $\omega_q = \Delta_B$ for qubit B. Each set comprises the average of 50 individual decay traces acquired sequentially in four-minute intervals. Set 1 exhibits purely exponential decay with $T_1 = 55\ \mu s$, whereas set 2 (acquired 17 hours after set 1) exhibits a non-exponential decay function. The black line is a fit to Eq. (3) assuming the non-exponential decay is due to quasiparticle fluctuations (see text). Inset: tabulation of the values obtained from fitting functions. The * indicates an assumed value from set 2 (not a fit value).

**(b)** Histograms of $T_1$ values obtained by exponential fits of the individual traces forming the two data sets in panel (a). For set 2, the fitting is restricted to the first $40\ \mu s$ to capture primarily the fast initial decay.

**(c)** Energy relaxation measurements at $\omega_q = \Delta_C$ for qubit C. The exponential decay function is manifest as a linear fit on the log plot with time constant $T_1 = 0.49~\mu s$.

**(d)** Histograms of $T_1$ values obtained from repeated measurements of qubit C. Both the exponential decay function (panel c) and the consistently tight $T_1$ distribution (panel d) indicate a relative insensitivity to quasiparticle number fluctuations.

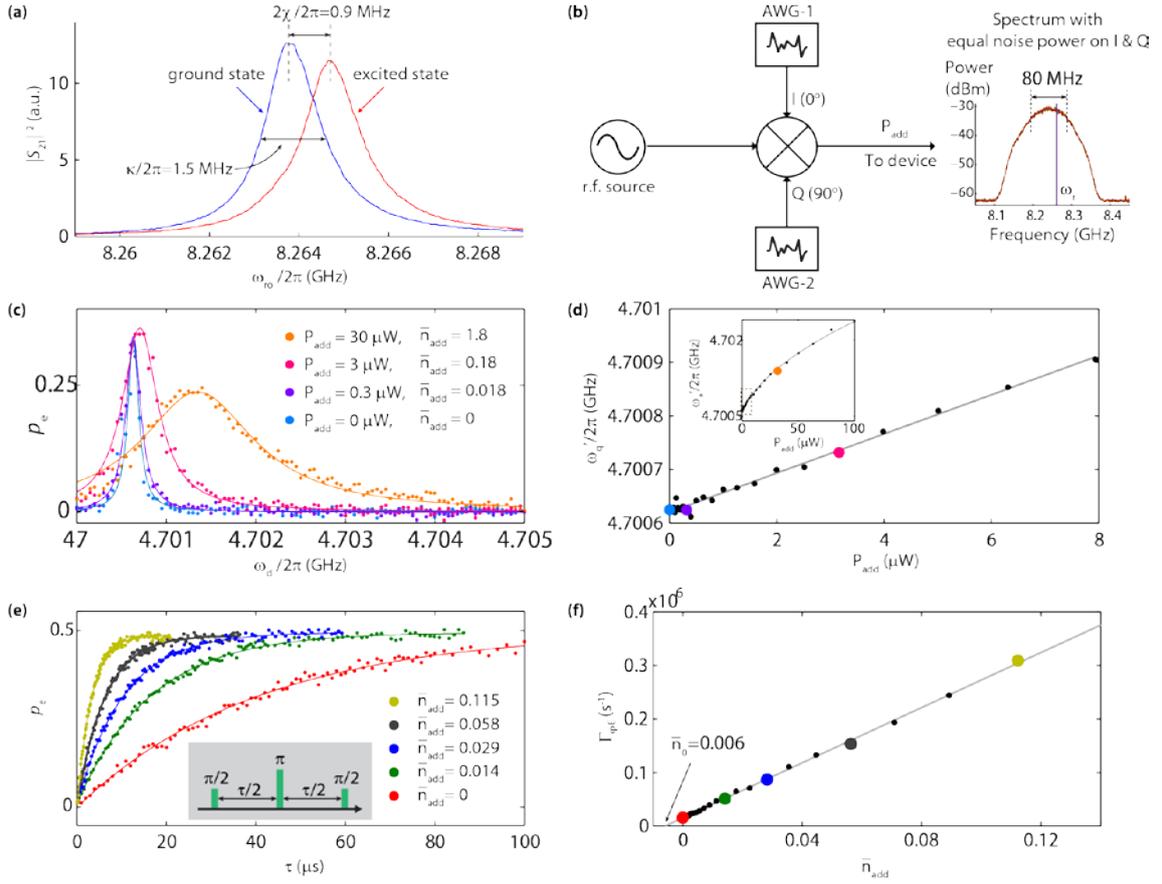

**Figure 5 | Calibration of engineered noise.**

**(a)** Resonator transmission spectra measured with the qubit prepared in the ground and excited states. In contrast to typical transmon qubits with $\omega_q < \omega_r$, an excited-state C-shunt flux qubit shifts the resonator to higher frequencies because of interactions with higher-level qubit transitions.

**(b)** Engineered thermal photon noise source. A coherent tone near the resonator frequency is mixed with white noise of nominally equal power from two independent arbitrary waveform generators (AWGs) applied to the in-phase (I) and quadrature (Q) ports of the I/Q mixer. The AWG noise bandwidth (80 MHz) is much greater than the cavity linewidth, creating effectively a thermal photon noise source with power $P_{\text{add}}$.

**(c)** Qubit spectral line shape (dots) and lorentzian fits (solid lines) for various added noise powers $P_{add}$. The equivalent photon population $\bar{n}_{add}$ added to the resonator is derived from (d). The blue trace corresponds to no added noise from the source in (b).

**(d)** Stark-shifted qubit frequency vs. applied noise power (dots). Colored dots correspond to traces in (c). Combining the linear fit (solid line) with the first-order dependence of Stark shift on photon population (Eq. 4) yields the power-per-added-photon $d\bar{n}_{add}/dP_{add} = 1/(17~\mu W)$ in the low-power limit. Inset: wider range of applied noise powers; dashed box indicates the range in the main panel. At large photon populations ($\bar{n}_{add} > 1$) the frequency shift becomes nonlinear, following Eq. (43) in Ref. 40 (solid line).

**(e)** Spin-echo decay (dots) with exponential fit (solid lines) for several values of added photons. Inset: spin-echo pulse sequence.

**(f)** Spin-echo pure-dephasing rate (echo decay rate without the $T_1$ contribution) plotted vs. injected photon population (dots). The linear fit (solid line) has slope $d\Gamma_{\varphi E}/d\bar{n}_{add} = 2.6 \times 10^6~\text{s}^{-1}$, in agreement with the value of $2.5 \times 10^6~\text{s}^{-1}$ calculated from Eq. (5). The intercept indicates a residual photon population $\bar{n}_0 = 0.006$ in the resonator.

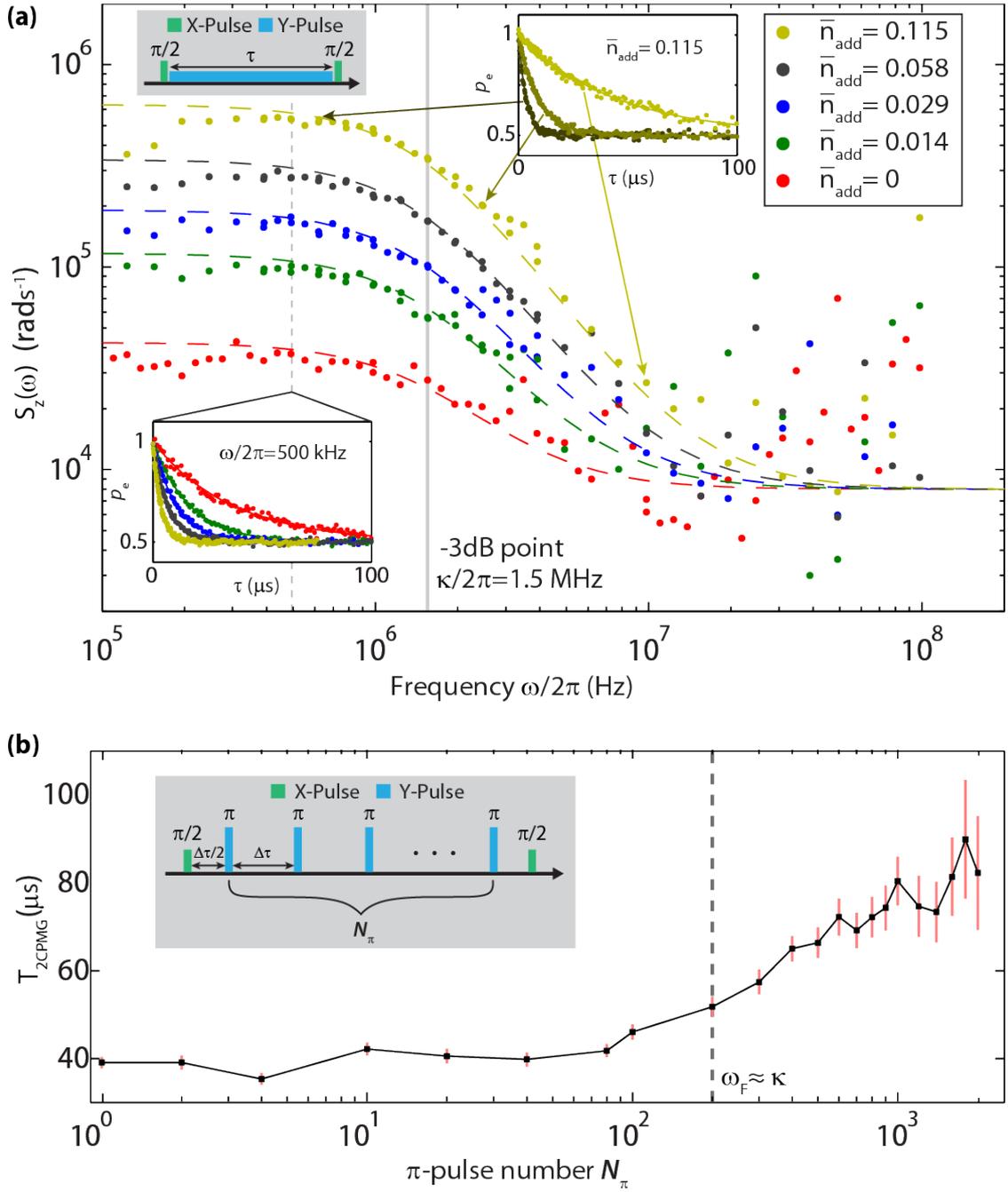

**Figure 6 | PSD of photon fluctuations in the resonator.**

(a) Noise power spectral densities (PSDs) extracted from spin-locking ($T_{1\rho}$) relaxation experiments (Inset, top-left: $T_{1\rho}$ pulse sequence) measured for different locking (Rabi) frequencies (0.1−100 MHz) and added noise photons $\bar{n}_{\text{add}}$. Colored dashed lines indicate expected lorentzian noise spectra [see Eq. (6)] assuming a constant white-noise offset. The vertical grey line indicates the 3-dB point of the lorentzians, coinciding with the resonator decay rate $\kappa/2\pi = 1.5$ MHz. Inset, bottom-left: $T_{1\rho}$ decay traces for different photon populations at fixed locking (Rabi) frequency $\omega/2\pi = \Omega_R/2\pi = 500$ kHz. Inset, top-right: $T_{1\rho}$ decay traces at different locking (Rabi) frequencies for $n_{add} = 0.115$.

**(b)** Decay times $T_{2CPMG}$ for the Carr-Purcell-Meiboom-Gill (CPMG) sequence measured vs. number of $\pi$-pulses, $N_\pi$, with $n_{add} = 0$ (no added noise photons). The CPMG pulse sequence (inset) acts as a band-pass noise filter centred at a frequency proportional to $N_\pi$ through the pulse spacing $\Delta\tau$ (see main text). At $N_\pi = 200$, the filter frequency approximately equals the cavity decay rate (dashed line). For $N_\pi < 200$, the filter samples the flat low-frequency portion of the lorentzian PSD, yielding a constant decay time $T_{2CPMG} \approx 40$ $\mu$s. For $N_\pi > 200$, the filter traverses the roll-off region of the lorentzian. As the sampled noise decreases, the decay times increase, approaching the limit set by energy relaxation ($T_{2CPMG} \approx 2T_1$) for $N_\pi > 1000$.

Supplementary Materials:

# The Flux Qubit Revisited to Enhance Coherence and Reproducibility

F. Yan[1], S. Gustavsson[1], A. Kamal[1], J. Birenbaum[2,*], A. Sears[3], D. Hover[3],

T. Gudmundsen[3], D. Rosenberg[3], G. Samach[3], S. Weber[3], J. Yoder[3],

T.P. Orlando[1,4], J. Clarke[2], A.J. Kerman[3], & W.D. Oliver[1,3,5]

[1]Research Laboratory of Electronics,
Massachusetts Institute of Technology, Cambridge, MA 02139, USA.
[2]Department of Physics, University of California, Berkeley, CA 94720-7300, USA.
[3]MIT Lincoln Laboratory, 244 Wood Street, Lexington, MA 02420, USA.
[4]Department of Electrical Engineering and Computer Science,
[5]Department of Physics,
Massachusetts Institute of Technology, Cambridge, MA 02139, USA.
*Current address: MIT Lincoln Laboratory; jeffrey.birenbaum@ll.mit.edu

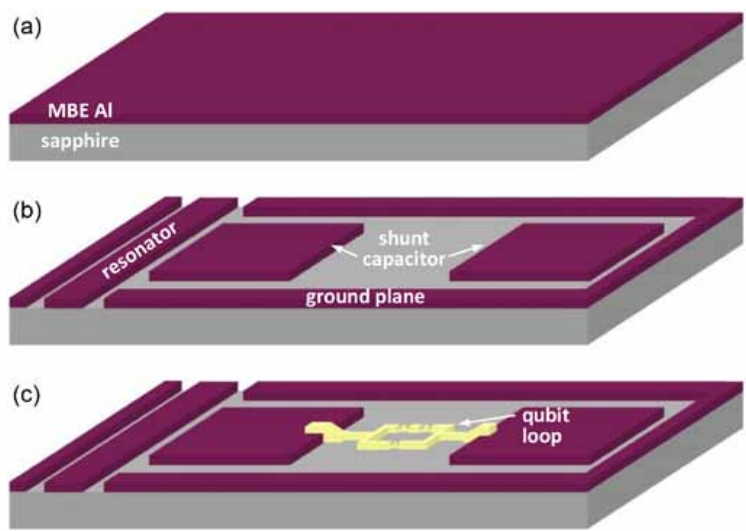

Supplementary Figure 1: **Process flow schematics of key steps for capacitively-shunted flux qubit fabrication. a**, Preparation of MBE aluminum (red) on outgassed C-plane sapphire substrates (gray). **b**, Patterning of the MBE aluminum into the shunt capacitor (representative square shunt capacitor geometry is shown), resonator center line, and surrounding ground plane. **c**, Patterning of the aluminum qubit loop (yellow), which contains three aluminum Josephson junctions. The loop contacts the shunt capacitor as illustrated.

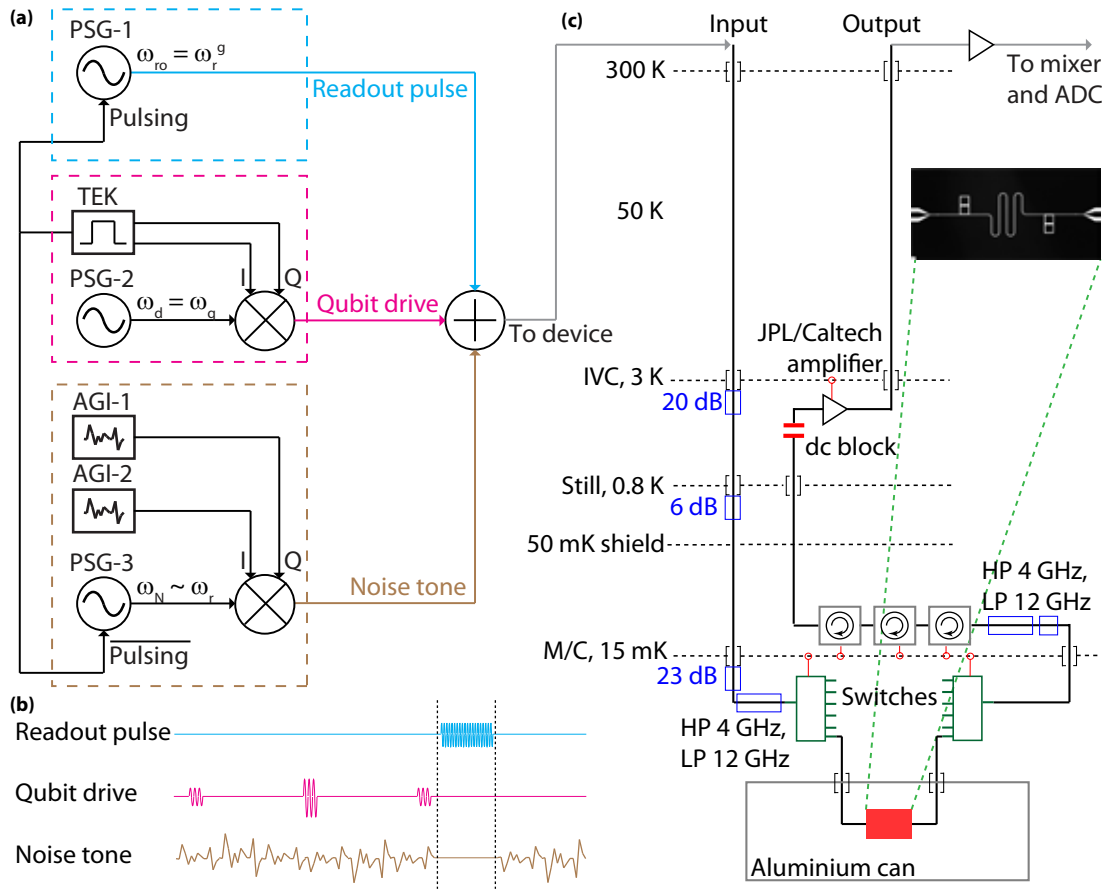

Supplementary Figure 2: **Experimental measurement schematic. a.** Diagram of the input-signal generation of readout (top panel, cyan), qubit drive (mid panel, magenta) and noise (bottom panel, brown) at room temperature. The signals are combined before being sent to the fridge. **b.** Relative timing of the signals generated in **a**. The dashed lines indicate the window within which the readout pulse is turned on while the noise is turned off. The representative qubit drive is a spin-echo pulse sequence, two $\pi/2$-pulses and a midpoint $\pi$-pulse. The actual pulses assume a Gaussian envelope. **c.** Wiring inside the refrigerator.

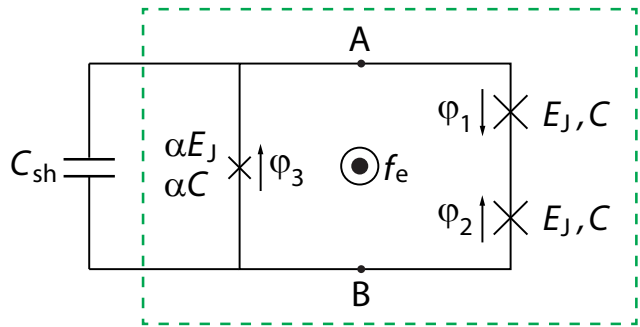

Supplementary Figure 3: **Circuit diagram of the C-shunt flux qubit.** Josephson junction 3 is $\alpha$ times the size of junctions 1 and 2. The external flux $f_e$ is defined as the magnetic flux threading the loop formed by the three junctions. Nodes A and B represent the superconducting islands. The shunt capacitor has the dominant capacitance in the circuit. Node A is also capacitively coupled to a superconducting resonator (not drawn). The green dashed box highlights the part of the circuit resembling the conventional persistent-current flux qubit.

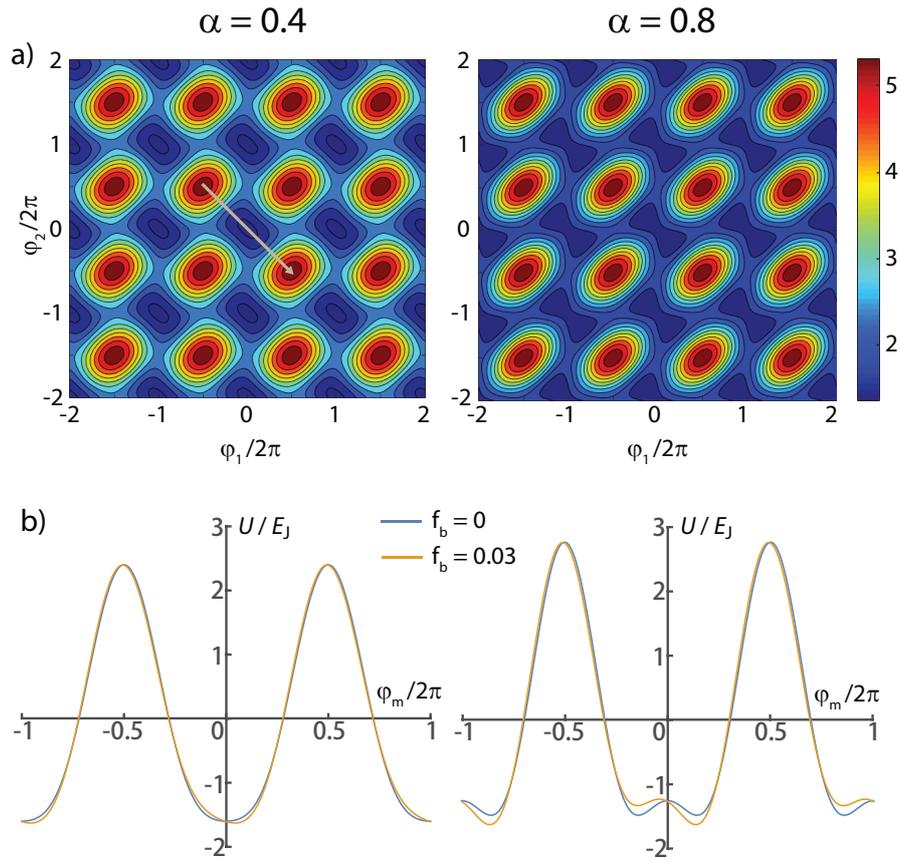

Supplementary Figure 4: **Potential profile of the C-shunt flux qubit. a.** A contour plot of the potential energy $U(\varphi_1, \varphi_2)$ at $f_b = 0$ for $\alpha = 0.4$ (C-shunt flux qubit, on the left) and $\alpha = 0.8$ (traditional flux qubit, on the right). In the C-shunt case, the square-shaped area centered around $(\varphi_1, \varphi_2) = (k_1 \cdot 2\pi, k_2 \cdot 2\pi)$ marks the single well. In the traditional case, the figure-eight-shaped area marks the double well. The gray arrow indicates the $\varphi_m = \varphi_1 - \varphi_2$ direction. **b.** The potential function along the $\varphi_m$ direction at $\varphi_1 + \varphi_2 = 0$ at $f_b = 0$ and $f_b = 0.03$ for both $\alpha = 0.4$ and $\alpha = 0.8$ cases.

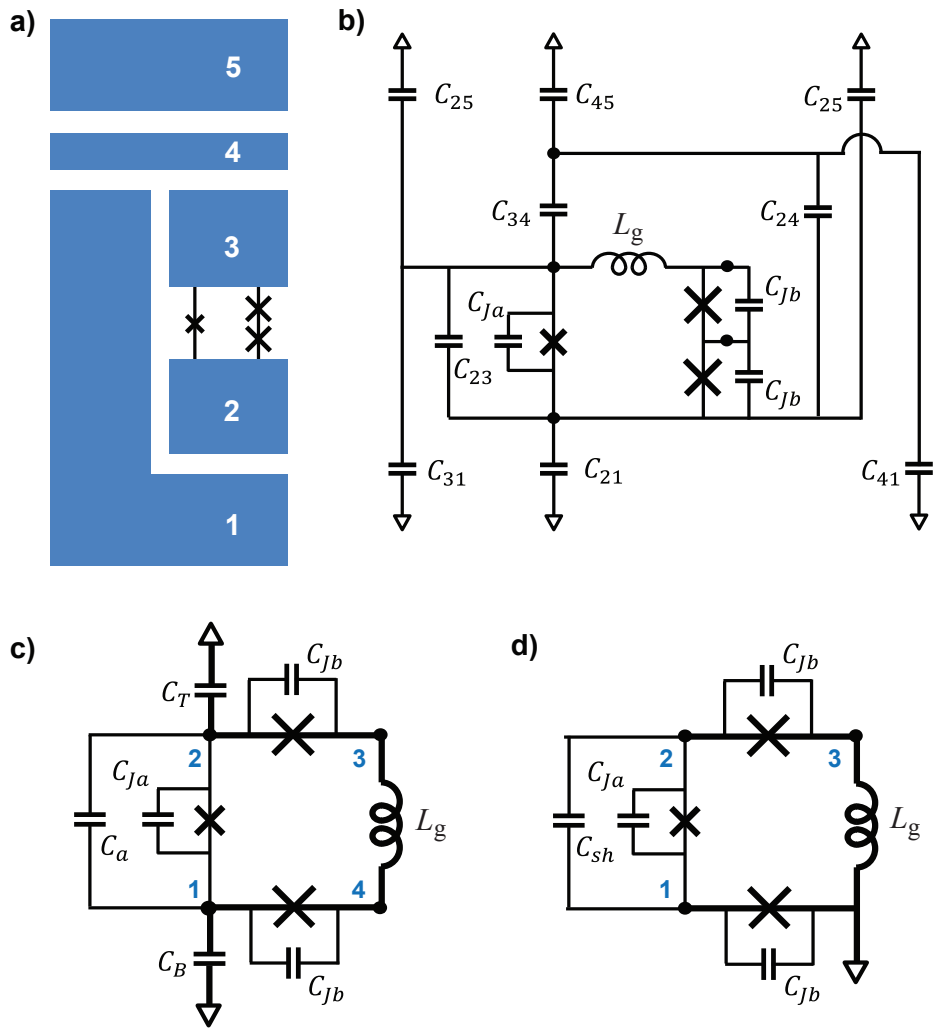

Supplementary Figure 5: **Circuit schematic for the C-shunt flux qubit and its reduction for use in simulation. a.** Schematic of the qubit-resonator system and the islands / nodes used in the simulation. For clarity the labeling follows that used for cQED transmons in Ref. 1. **b.** Equivalent circuit schematic with islands 1 and 5 grounded. Filled black circles indicate the five nodes that will be used in the full simulation. **c.** First reduction of the circuit in (b) containing four nodes (labeled in blue). See text for details. **d.** Second reduction of the circuit in (b) containing three nodes (labeled in blue). An effective shunt capacitance $C_\mathrm{sh}$ accounts for the capacitances $C_\mathrm{T}$, $C_\mathrm{B}$ and $C_\mathrm{a}$. See text for details.

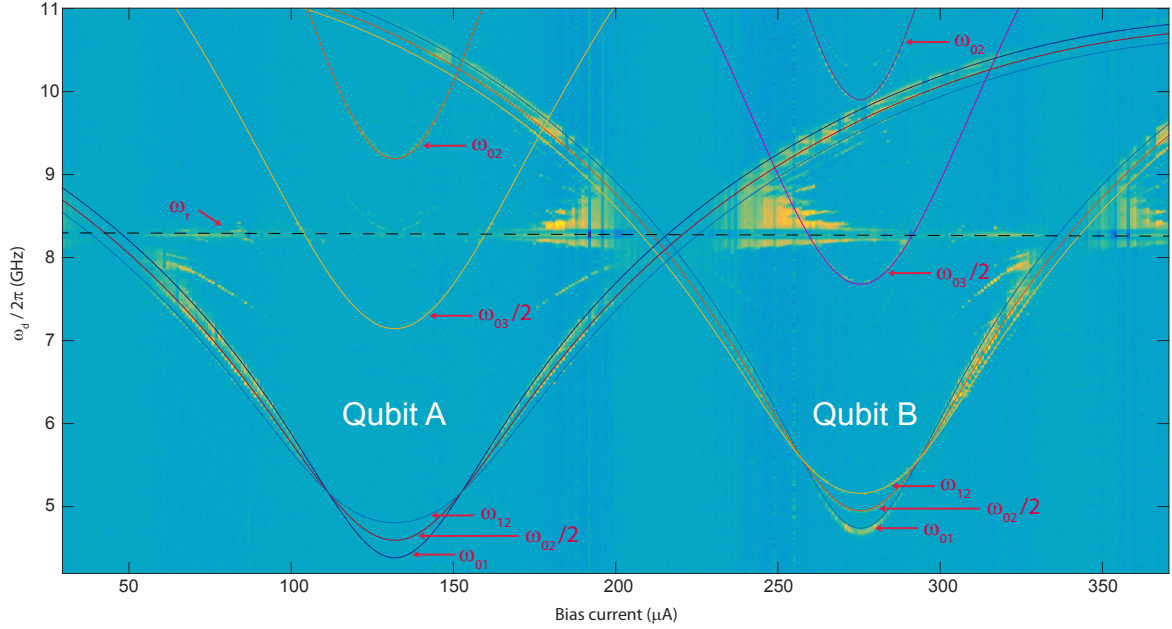

Supplementary Figure 6: **High-power spectra of qubits A and B with simulations to match the higher-level transitions.** The horizontal axis corresponds to the current in the "bobbin" coil of wire local to the qubits to apply a magnetic flux bias. Solid lines are simulation results that give the best match to the visible 0–1, 0–2 (two photon), 1–2, 0–3 (two photon) and 0–2 transitions. A precise matching to the experimental spectra required allowing these parameters to vary somewhat from their design values (Supplementary Table 1). Nonetheless, the simulation parameters (*e.g.*, $J_c$, $C_{sh}$, *etc.*) are generally same for the two qubits, with the primary exception being the junction sizes.

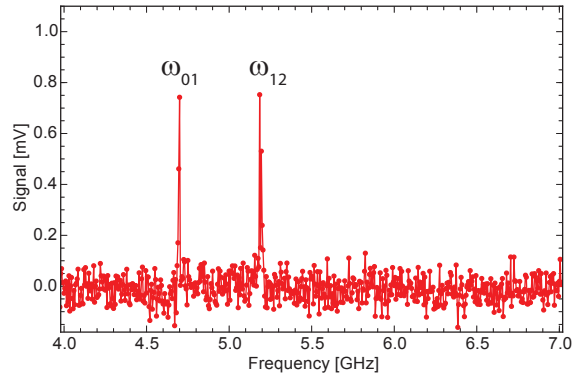

Supplementary Figure 7: **Two-tone spectroscopy of qubit B measured at the optimal bias point ($\Phi_e = \Phi_0/2$).** The data are taken by first applying a 20 ns microwave pulse at $\omega/2\pi = \omega_{01}/2\pi = 4.701$ GHz to drive the qubit to the $|1\rangle$ state, followed by a long (50 $\mu$s) low-power microwave tone. The frequency of the second tone is swept to perform spectroscopy. The resulting spectrum shows two peaks, one at 4.701 GHz that corresponds to driving the 1–0 transition, and one at 5.191 GHz that corresponds to the 1–2 transition. The measured anharmonicity is $(\omega_{12} - \omega_{01})/2\pi = 490$ MHz.

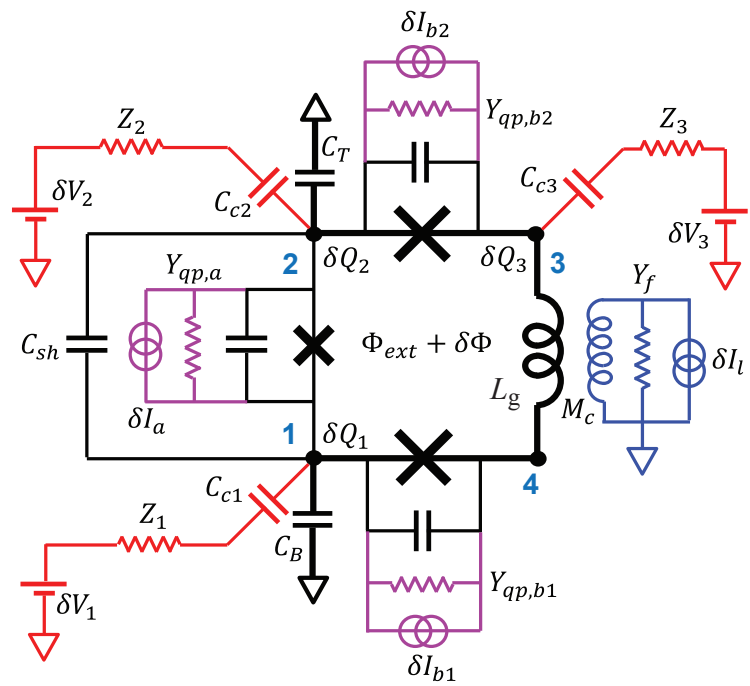

Supplementary Figure 8: **Circuit schematic used to account for electric and magnetic noise coupled to the qubit.** See main text Supplementary Note Supplementary Note 8 for details.

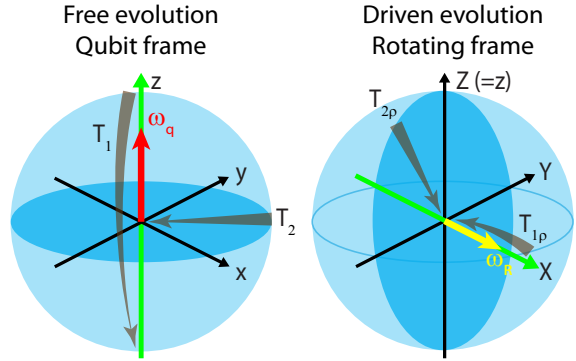

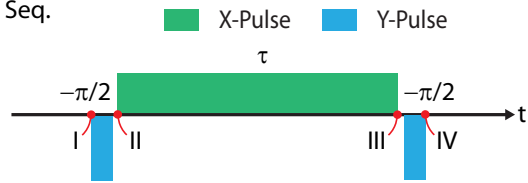

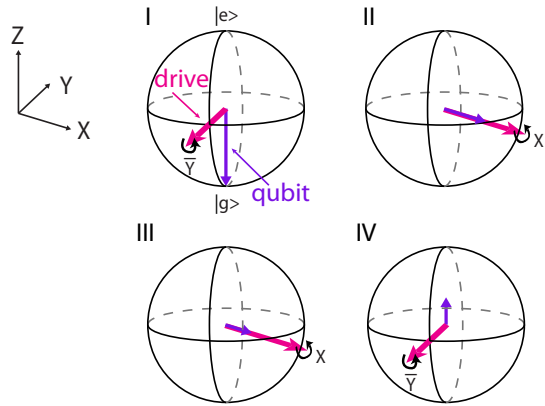

Supplementary Figure 9: **Spin locking and its implementation for noise spectroscopy. a.** Analogy between free- (left) and driven-evolution (right) dynamics. The free evolution is described in the qubit frame $\{x, y, z\}$ while the driven evolution is described in the rotating frame $\{X, Y, Z\}$. The two cases differ in the orientation and size of the static field, *i.e.,* "$\omega_q \sigma_z/2$" (red arrow) versus "$\omega_R \sigma_X/2$" (yellow arrow). The corresponding quantization axes (green arrows) and longitudinal/transverse relaxations (gray arrows) are defined with respect to the static field. **b.** Standard three-pulse spin-locking sequence (SL-3). The long driving pulse is $90°$-phase-shifted from the $\pi/2$-pulses, and its length $\tau$ is the scanned parameter to record the rotating-frame relaxation. **c.** Bloch sphere representation of the rotating-frame qubit dynamics under SL-3. The purple arrows represent the polarization of the qubit states, while the magenta arrows indicate the driving-field orientation. The qubit is initially prepared in its ground state (I). The first $\pi/2$-pulse rotates the qubit by $90°$ into the equatorial plane (II). The second $90°$-phase-shifted continuous driving pulse, of duration $\tau$, is then aligned with the qubit state, effectively locking the qubit along X. During the pulse, the qubit undergoes relaxation in this rotating frame towards its steady state (III). The final $\pi/2$-pulse projects the remaining polarization onto Z (=z) for readout (IV).

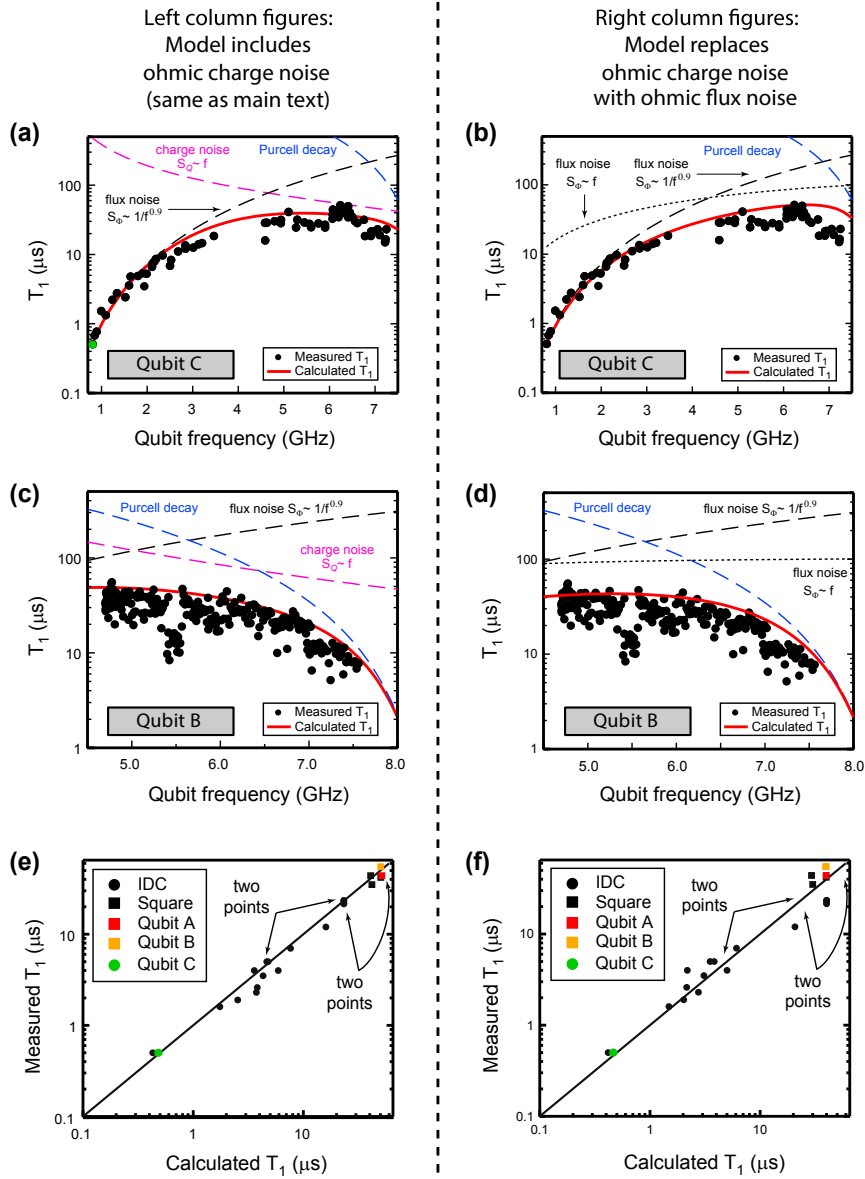

Supplementary Figure 10: **Comparison of a model using ohmic charge noise with one using ohmic flux noise.** Panels **a**, **c**, and **e** reproduce the panels in Fig. 3 from the main text, and the simulation traces contained therein are derived from a model involving ohmic charge noise, $1/f$ flux noise and the Purcell effect. For comparison, panels **b**, **d**, and **f** replace the ohmic charge noise with ohmic flux noise. The agreement with the data is similar between the two models, making it challenging to distinguish between ohmic charge noise and ohmic flux noise. It is only in panels **a** and **b** (device C) at frequencies above 6 GHz where the qubit is especially sensitive to ohmic charge noise that a plausible distinction can be observed.

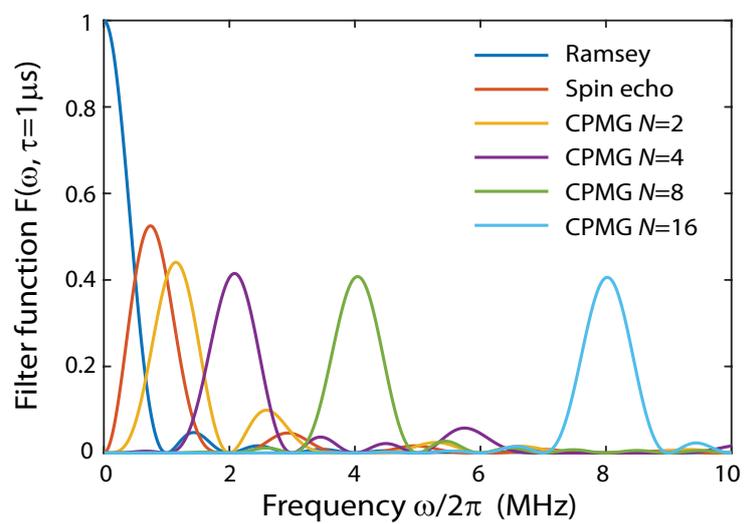

Supplementary Figure 11: **System filter functions corresponding to representative control sequences.** Filter functions of Ramsey, spin-echo and CPMG ($N = 2, 4, 8, 16$) sequences, assuming $\tau = 1\,\mu$s in all cases.

| Sample | | Shunt-cap. and Coupling | | | | Junction Parameters | | | | | Qubit Parameters at $\Phi_b = 0$ | | | | | | Measured | |
|---|---|---|---|---|---|---|---|---|---|---|---|---|---|---|---|---|---|---|
| # | main text | cap type | gap ($\mu$m) | $C_{sh}$ (fF) | $\beta_c$ | $\alpha$ | $J_c$ ($\mu$A/$\mu$m$^2$) | $E_{J\alpha}$ (GHz) | $E_{C\alpha}$ (GHz) | $\Delta$ (GHz) | $\mathcal{A}$ (GHz) | $V_{eg}$ ($\mu$V) | $I_p$ (nA) | $T_1^\Phi$ ($\mu$s) | $T_1^Q$ ($\mu$s) | $T_1^P$ ($\mu$s) | $T_1^{sim}$ ($\mu$s) | $T_1^{meas}$ ($\mu$s) | $\Delta$ (GHz) |
| 1 | | IDC | 40 | 21.3 | 0.19 | 0.55 | 6.7 | 139 | 0.74 | 3.36 | 6.15 | 4.28 | 194 | 4.4 | 21 | 742 | 3.6 | 4.0 | 4.4 |
| 2 | | IDC | 40 | 21.3 | 0.19 | 0.55 | 6.7 | 139 | 0.74 | 3.36 | 6.15 | 4.28 | 194 | 4.8 | 19 | 742 | 3.8 | 2.6 | 4.9 |
| 3 | | IDC | 20 | 21.3 | 0.19 | 0.59 | 6.6 | 127 | 0.75 | 1.50 | 8.37 | 2.46 | 206 | 1.8 | 155 | 3014 | 1.8 | 1.6 | 1.8 |
| 4 | | IDC | 20 | 17.5 | 0.19 | 0.63 | 6.6 | 137 | 0.88 | 0.88 | 11.65 | 1.66 | 275 | 0.5 | 821 | 2894 | 0.5 | 0.5 | 0.7 |
| 5 | C | IDC | 40 | 9.0 | 0.31 | 0.68 | 6.6 | 133 | 1.46 | 0.82 | 17.8 | 1.76 | 275 | 0.5 | 685 | 2894 | 0.5 | 0.5 | 0.8 |
| 6 | | IDC | 40 | 21.3 | 0.19 | 0.55 | 6.6 | 137 | 0.74 | 3.36 | 6.10 | 4.30 | 191 | 2.7 | 37 | 742 | 2.5 | 1.9 | 2.5 |
| 7 | | IDC | 40 | 22.7 | 0.31 | 0.53 | 5.9 | 93.1 | 0.73 | 3.88 | 4.40 | 4.74 | 126 | 8.4 | 22 | 241 | 5.9 | 4.0 | 3.5 |
| 8 | | IDC | 20 | 27.6 | 0.21 | 0.52 | 5.9 | 87.3 | 0.62 | 3.94 | 3.51 | 4.45 | 113 | 15 | 16 | 408 | 7.6 | 7.0 | 5.4 |
| 9 | | IDC | 20 | 17.9 | 0.23 | 0.56 | 5.9 | 105 | 0.88 | 3.31 | 6.29 | 4.61 | 152 | 6.3 | 21 | 536 | 4.8 | 5.0 | 3.9 |
| 10 | | IDC | 20 | 17.9 | 0.23 | 0.55 | 5.9 | 105 | 0.88 | 4.19 | 5.53 | 5.34 | 145 | 6.7 | 16 | 305 | 4.7 | 5.0 | 3.7 |
| 11 | | IDC | 40 | 21.3 | 0.19 | 0.55 | 5.9 | 122 | 0.74 | 3.35 | 5.75 | 4.30 | 171 | 4.4 | 28 | 762 | 3.7 | 2.3 | 3.3 |
| 12 | | IDC | 40 | 21.3 | 0.19 | 0.54 | 5.9 | 122 | 0.74 | 4.14 | 5.05 | 4.90 | 163 | 5.8 | 18 | 467 | 4.3 | 3.5 | 4.1 |
| 13 | | IDC | 20 | 26.8 | 0.15 | 0.52 | 3.0 | 44.3 | 0.64 | 3.19 | 2.73 | 4.05 | 61.8 | 32 | 33 | 1378 | 16 | 12 | 3.2 |
| 14 | | IDC-R | 10 | 51.4 | 0.14 | 0.43 | 2.4 | 36.2 | 0.35 | 3.94 | 0.88 | 3.43 | 43.9 | 82 | 34 | 643 | 23 | 23 | 4.2 |
| 15 | | IDC-R | 10 | 51.4 | 0.14 | 0.43 | 2.4 | 36.2 | 0.35 | 3.94 | 0.88 | 3.43 | 43.9 | 82 | 34 | 643 | 23 | 22 | 4.2 |
| 16 | | IDC-R | 30 | 51.9 | 0.14 | 0.43 | 2.4 | 36.2 | 0.35 | 3.94 | 0.87 | 3.43 | 43.9 | 82 | 34 | 643 | 23 | 24 | 4.2 |
| 17 | | IDC-R | 30 | 51.9 | 0.14 | 0.43 | 2.4 | 36.2 | 0.35 | 3.94 | 0.87 | 3.43 | 43.9 | 82 | 34 | 643 | 23 | 22 | 4.2 |
| 18 | | SQR | 40 | 51.0 | 0.089 | 0.43 | 2.9 | 43.0 | 0.35 | 4.33 | 0.88 | 3.60 | 50.3 | 68 | 147 | 408 | 42 | 35 | 4.7 |
| 19 | | SQR | 40 | 51.0 | 0.089 | 0.43 | 2.9 | 43.0 | 0.35 | 4.68 | 0.88 | 3.74 | 50.1 | 75 | 124 | 330 | 41 | 44 | 5.1 |
| 20 | | SQR | 40 | 51.0 | 0.089 | 0.42 | 2.4 | 36.2 | 0.35 | 3.96 | 0.80 | 3.45 | 43.9 | 79 | 184 | 567 | 50 | 42 | 4.0 |
| 21 | A | SQR | 40 | 51.0 | 0.089 | 0.43 | 2.4 | 36.2 | 0.35 | 3.96 | 0.91 | 3.45 | 43.9 | 85 | 169 | 567 | 52 | 44 | 4.4 |
| 22 | B | SQR | 40 | 51.0 | 0.089 | 0.42 | 2.4 | 36.2 | 0.35 | 4.29 | 0.83 | 3.60 | 43.5 | 92 | 145 | 474 | 50 | 55 | 4.7 |

Supplementary Table 1: Parameters for 22 qubits studied in this work, spanning shunt capacitance values $C_{sh} = 9\ldots 52$ fF. Color-highlighting corresponds to the samples A, B and C presented in detail in the main text. All parameters are design values or are derived from simualtion, except for the critical current density $J_c$ and the measured relaxation time $T_1^{meas}$. Three capacitor types were studied: IDC is "interdigital capacitor"; IDC-R is an IDC with slightly rounded corners (negligible when compared with actualized standard IDCs, but included here for completeness); and SQR is square capacitor shape. Gap is the spacing between capacitor features (*e.g.,* IDC fingers). $\beta_c$ quantifies the coupling strength between the qubit and the resonator. $E_{J\alpha}$ and $E_{C\alpha}$ are respectively the Josephson energy and charging energy of the small junction. $\Delta$ is the qubit frequency at the qubit flux-insensitive (degeneracy) point $\Phi_b = 0$. $\mathcal{A}$ is the anharmonicity. $V_{eg}$ is the simulated island node voltage (related to the charge transverse matrix element). $I_p$ is the simulated persistent current (related to the flux transverse matrix element). $T_1^\Phi$, $T_1^Q$, and $T_1^P$ are respectively the simulated $T_1$ values due to flux noise, charge noise, and the Purcell effect (all at $\Phi_b = 0$). $T_1^{sim}$ is the net simulated $T_1$ value due to these processes, and and $T_1^{meas}$ is the measured relaxation time at $\Phi_b = 0$.

# Supplementary Note 1   Materials and Fabrication of the High-Q Capacitor

The capacitively-shunted flux qubits studied in this work were prepared using the following steps:

1. Growth and patterning of high-quality-factor (high-Q) aluminum films using molecular beam epitaxy (MBE).

2. Patterning and evaporation of the superconducting qubit loop and Josephson junctions.

3. Dicing and packaging.

## Supplementary Note 1.1   Growth and patterning of high-Q aluminum

High-Q aluminum films were deposited on 50-mm C-plane sapphire wafers in a Veeco GEN200 MBE system with a growth chamber base pressure of $10^{-11}$ torr. The wafers were cleaned in piranha solution (sulfuric acid and hydrogen peroxide) prior to loading into the MBE system. The wafers were annealed in the MBE system at $900\,°C$ to facilitate outgassing and sapphire surface reconstruction, after which 250 nm of aluminum was deposited at a growth rate of 0.025 nm/s and a substrate temperature of $150\,°C$ (Supplementary Figure 1a).

The high-Q aluminum was patterned using contact lithography and wet-etched using Aluminum Etchant - Type A (Transene Company, Inc.) into the following device features (Supplementary Figure 1b): shunt capacitors, coplanar waveguide (CPW) resonators, ground planes, and optical alignment marks.

## Supplementary Note 1.2   Patterning the qubit loop and Josephson junctions

The qubit loop and junctions were formed using double-angle evaporation of aluminum through Dolan-style bridges (Supplementary Figure 1c) [2]. The free-standing bridges were realized using a bilayer mask comprising a germanium hard mask on top of a sacrificial MMA/MAA layer [MicroChem methyl methacrylate (MMA (8.5)/MAA EL9)]. The qubit loop and junctions were patterned using electron-beam lithography (Vistec EBPG5200) and ZEP520A resist (ZEONREX Electronic Chemicals). This pattern was transferred into the Ge layer using a $CF_4$ plasma, and the underlying MMA/MAA resist was under-etched using an oxygen plasma to create free-standing bridges. Prior to the aluminum evaporation, an *in situ* argon ion milling was used to clean exposed contact points on the MBE aluminum to ensure superconducting contact with the evaporated aluminum. The qubit loops and junctions were realized with two separate angle-evaporated aluminum layers; between the two aluminum evaporation steps, static oxidation conditions were used to prepare junctions with a certain critical current density.

## Supplementary Note 1.3   Dicing and packaging

Devices were diced into 2.5 x 5 mm$^2$ chips (as shown in the manuscript in Fig. 1a) that were mounted into gold-plated copper packages. Aluminum wirebonds were used for both signal and ground connections between the device and package, as well as to connect the ground planes of the CPW resonator to prevent slotline modes.

## Supplementary Note 2  Measurement Set-up and Protocol

We performed our experiments at MIT in a Leiden Cryogenics (CF-450) dilution refrigerator with a base temperature of 15 mK. The device was magnetically shielded with a superconducting can surrounded by a Cryoperm-10 cylinder. All electrical leads were attenuated and/or filtered to reduce noise.

### Supplementary Note 2.1  Outside the dilution refrigerator

The electronic setup for generating the readout-pulses, control-pulses, and artificial photon noise is shown in Supplementary Figure 2a. All time-sensitive instruments are synchronized with a Stanford Research Systems FS725 Rubidium Frequency Standard. The readout-pulse is generated by an Agilent 8267D Vector Signal Generator (PSG), gated by a Tektronix 5014b arbitrary waveform generator (TEK). This pulse is typically a few microseconds long, and the tone is set at the resonator frequency dressed by the qubit in its ground state ($\omega_r^g/2\pi$). The control-pulse envelope for driving the qubit is generated by the TEK and mixed with a qubit-frequency tone ($\omega_q/2\pi$) from a second PSG using its internal I-Q mixer. These pulses are further gated to reduce unwanted mixer leakage. Artificial photon noise is generated by up-converting (quadrature-mixing) 80-megahertz-wide white-noise signals from two Agilent 33250A arbitrary waveform generators (AGI) with a tone near the cavity frequency from a third PSG. We confirmed that the noise source behaved consistently using either one port or both ports of the I-Q mixer. For example, with equal powers applied to each port, the output is simply a doubling of the photon noise power generated from a single port. The carrier frequency for the noise is chosen to be ~10 MHz away from the readout frequency, much larger than the cavity linewidth. In addition, we gate the noise off during the readout pulse using the TEK. The three signals are combined (relative timing illustrated in Supplementary Figure 2b) and sent to the input port on the refrigerator. This sequence constitutes a single experimental trial, and it is repeated typically 10,000 times with a period of 200-400 $\mu$s.

### Supplementary Note 2.2  Inside the dilution refrigerator

The device is enclosed in a copper package, which is itself mounted inside an aluminum box to shield the device from external electromagnetic radiation and magnetic field fluctuations (Supplementary Figure 2c). A small coil antenna is mounted under the package lid and is used to bias the qubit with a static magnetic field (not shown). A Yokogawa 7651 dc source provides the bias current to the coil using twisted pair wires passing through an RC low-pass filter with cutoff around 100 kHz. On the input side, there is a total attenuation of 49 dB arising from discrete attenuators (XMA Corporation) at various temperature stages. In addition, there is a relatively small amount of distributed attenuation due to loss in the coaxial cables. On both the input and output sides, a high-pass filter (RLC F-18948, 4 GHz cutoff) and a low-pass filter (RLC L-3615, 12.4 GHz cutoff) provide a net 4-12.4 GHz passband. After the output filters, 3 isolators (Quinstar / Pamtech, model CWJ1019KS414, 3-12 GHz, with approximately 15-20 dB isolation each) are mounted on the mixing chamber. The output signal is amplified by a JPL/Caltech cryogenic preamplifier (1-12 GHz, 30 dB gain). The output port outside the refrigerator is followed by a room-temperature amplifier (MITEQ, AMF-5D-00101200-23-10P, 0.1-12 GHz, 43 dB gain) before mixing with a local oscillator (LO) tone ($\omega_{ro}/2\pi - \omega_{LO}/2\pi = 50$ MHz) for heterodyne detection (not shown). After the mixers, the signal is digitized using an Acquiris U1084A analog-to-digital converter (ADC) and digitally demodulated to extract the amplitude and phase of the readout signal for qubit-state estimation.

# Supplementary Note 3    Two-Level-System Model of C-Shunt Flux Qubit

## Supplementary Note 3.1    Parameterization of the two-level-system Hamiltonian

In the main text, we elect to parameterize the C-shunt flux qubit using the familiar notation of the conventional flux qubit [3], albeit with important generalizations due to the influence of higher energy levels. In the laboratory-frame, the two-level-system Hamiltonian for an individual C-shunt flux qubit near flux-degeneracy and coupled to a CPW resonator is:

$$\hat{\mathcal{H}} = \frac{\hbar}{2}\left[\Delta\left(\Phi_\text{b}\right)\hat{\sigma}_\text{x} + \varepsilon\left(\Phi_\text{b}\right)\hat{\sigma}_\text{z}\right] + \hbar\omega_\text{r}(\hat{a}^\dagger\hat{a} + 1/2) + \hbar g\left(\Phi_\text{b}\right)\hat{\sigma}_\text{y}(\hat{a}^\dagger + \hat{a}). \tag{1}$$

Here, the three terms are respectively the qubit, resonator, and qubit-resonator interaction Hamiltonians, $\hat{\sigma}_\text{x,y,z}$ are the Pauli operators for the qubit, $\Phi_\text{b} \equiv \Phi_\text{e} - \Phi_0/2$ is the flux bias due to an external magnetic flux $\Phi_\text{e}$ ($\Phi_0 \equiv h/2e$ is the superconducting flux quantum), $\hat{a}^\dagger(\hat{a})$ is the raising(lowering) operator for resonator photons, $\omega_\text{r}$ is the resonator angular frequency and $g$ is the qubit-resonator coupling angular frequency. In this coordinate system, the qubit shunt-capacitor couples transversally to the qubit through $\hat{\sigma}_\text{y}$. Similarly, charge fluctuations connect through $\hat{\sigma}_\text{y}$.

Within our parametrization, the circulating current states of the C-shunt flux qubit have energies $\pm\hbar\varepsilon(\Phi_\text{b})/2 \equiv \pm I_\text{m}(\Phi_\text{b})\Phi_\text{b}$, where $I_\text{m}(\Phi_\text{b}) \equiv [I_\text{p1}(\Phi_\text{b}) - I_\text{p0}(\Phi_\text{b})]/2$ is related to the difference in the flux-dependent circulating (persistent) currents $I_\text{p0}(\Phi_\text{b})$ and $I_\text{p1}(\Phi_\text{b})$, and these states hybridize with a flux-dependent energy $\hbar\Delta(\Phi_\text{b})$. With this generalization, a flux-dependent offset energy $\hbar\varepsilon_\text{off}(\Phi_\text{b}) \equiv I_\text{off}(\Phi_\text{b})\Phi_\text{b}$ is removed, where $I_\text{off}(\Phi_\text{b}) \equiv [I_\text{p1}(\Phi_\text{b}) + I_\text{p0}(\Phi_\text{b})]/2$. In addition to their flux dependence, the currents $I_\text{p0,p1}(\Phi_\text{b})$ generally do not have the same magnitude, nor need they have opposing sign. This is in marked contrast to the conventional flux qubit, where $I_\text{p0} = -I_\text{p1} \equiv I_\text{p}$ with both $I_\text{p}$ and $\Delta$ being independent of flux within a certain region about $\Phi_\text{b} = 0$. The fact that $I_\text{p0,p1}(\Phi_\text{b})$ and $\Delta(\Phi_\text{b})$ are flux-dependent for the C-shunt flux qubit reflects the non-negligible role of its higher energy levels. Consequently, we numerically diagonalize the full qubit-resonator Hamiltonian for parameter extraction and modeling.

The extent to which higher levels of the qubit influence the parameters in Eq. (1) is determined by the ratio $\omega_\text{p}^{(k)}/\omega_\text{q}$ where $\omega_\text{q} = \sqrt{\varepsilon(\Phi_\text{b})^2 + \Delta(\Phi_\text{b})^2}$ is the qubit frequency, and $\omega_\text{p}^{(k)}$ are the plasma frequencies of the additional oscillator-like modes of the circuit. For conventional flux qubits, these modes are typically much higher in energy than the qubit excited state and can be neglected. However, as the shunt capacitance is increased, at least one of these modes shifts to lower frequency and can eventually become comparable to the qubit frequency itself. For example, one of these modes is the small junction "plasma frequency" $\omega_{\text{p},\alpha} = 1/\sqrt{L_{\text{J},\alpha}C_\alpha}$ determined by its Josephson inductance $L_{\text{J},\alpha}$ and its total capacitance $C_\alpha$ (including the shunt capacitor). This is the frequency of the oscillator-like degree of freedom for the qubit in each of the two wells of its double-well potential profile (each well is associated with a circulating persistent current $I_\text{p0,p1}$), and this frequency is ideally far above $\omega_\text{q}$. For a conventional flux qubit, $\omega_\text{p}^{(k)}/\omega_\text{q} \gtrsim 10$, and the parameters in Eq. (1) are largely flux-independent over a wide range of flux bias about $\Phi_0/2$ (see Ref. 4 for experiments that study the flux range required to access higher levels for a particular conventional flux qubit). As the ratio $\omega_\text{p}^{(k)}/\omega_\text{q}$ is decreased, the excited states of the plasma mode shift closer to the two qubit levels and quantum mixing (hybridization) occurs, resulting in increased quantum fluctuations of the current. Correspondingly, there is a decrease in the range in flux over which the parameters in Eq. (1) are essentially flux independent. Eventually, when $\omega_\text{p}^{(k)}/\omega_\text{q} \gtrsim 1$, the qubit can be described as a weakly-harmonic oscillator (much like a transmon), where the only remnants of the persistent currents are small, state-dependent displacements of the oscillator current from zero, which are much smaller than their quantum fluctuations. Typically, the resulting weak anharmonicity is inverted relative to that of the transmon. The highest-coherence C-shunt flux qubits considered in this work are in an intermediate regime, having $\omega_\text{p}^{(k)}/\omega_\text{q} \gtrsim 2$. Their anharmonicity is around 500-900 MHz for the longest lived devices, generally larger than the 200-300 MHz observed in transmons.

The role of higher energy levels is described in Section Supplementary Note 4, along with a comparison between the conventional and C-shunt flux qubits. Ultimately, to account for the higher energy levels, the two-level parametrization must be deduced from experiment, an analytic treatment (Section Supplementary Note 4), or simulations (Section Supplementary Note 5).

## Supplementary Note 3.2 Dispersive two-level-system Hamiltonian

The unitary $\hat{U}_1 = \exp\left[i\left(\frac{\pi}{2} - \theta\right)\hat{\sigma}_y/2\right]$ rotates Eq. (1) from the laboratory frame to the qubit frame,

$$\hat{U}_1 \hat{\mathcal{H}} \hat{U}_1^\dagger = \hbar\omega_q \hat{\sigma}_z/2 + \hbar\omega_r (\hat{a}^\dagger \hat{a} + 1/2) + \hbar g \hat{\sigma}_y (\hat{a}^\dagger + \hat{a}) \,, \tag{2}$$

where $\theta = \arctan(\varepsilon/\Delta)$. The transverse qubit-resonator coupling is unaffected by the transformation, since the flux bias $\Phi_b$ rotates the qubit quantization axis within the x–z plane (laboratory frame) and the capacitive coupling is transverse to this plane.

In the dispersive regime, $|\delta_{q,r}| = |\omega_q - \omega_r| \gg g$, after a second unitary transformation $\hat{U}_2 = \exp\left[-i \frac{g}{\Delta_{q,r}} (\hat{a}\hat{\sigma}_+ + \hat{a}^\dagger \hat{\sigma}_-)\right]$, the Hamiltonian can be approximated:

$$\hat{\mathcal{H}}_{\text{disp}} = \hat{U}_2 \hat{U}_1 \hat{\mathcal{H}} \hat{U}_1^\dagger \hat{U}_2^\dagger \approx \hbar\omega_q \hat{\sigma}_z/2 + \hbar\omega_r (\hat{a}^\dagger \hat{a} + 1/2) + \hbar 2\chi (\hat{a}^\dagger \hat{a} + 1/2) \hat{\sigma}_z/2 \,. \tag{3}$$

Here, $\omega_q = \sqrt{\varepsilon(\Phi_b)^2 + \Delta(\Phi_b)^2}$ is the qubit angular frequency and $\chi(\Phi_b)$ is related to the qubit-state-dependent dispersive shift $2\chi(\Phi_b) = g^2(\Phi_b)/\delta_{q,r}(\Phi_b)$ of the resonator used for readout. The last term includes the Stark shift $\Delta_{\text{Stark}} = 2\chi(\Phi_b)\hat{n}$ due to the resonator photon number $\hat{n} = \hat{a}^\dagger \hat{a}$ and the Lamb shift $\Delta_{\text{Lamb}} = \chi(\Phi_b)$ due to the resonator zero-point energy. Higher levels of the qubit generally play an important role in constructing the flux-dependent dispersive shift. While it is straightforward to measure this experimentally, one must go beyond a two-level approximation to calculate it accurately.

# Supplementary Note 4 Analytic Treatment and Comparison of the Conventional and C-Shunt Flux Qubits

## Supplementary Note 4.1 Conventional persistent-current flux qubit

Before developing the system model for the C-shunt flux qubit, we review the conventional persistent-current flux qubit as presented in Ref. [3]. The conventional flux qubit circuit is illustrated in Supplementary Figure 3 (dashed box). The qubit loop is interrupted by three Josephson junctions, and $\varphi_i$ ($i = 1, 2, 3$) are the associated gauge-invariant phase differences. Two of the junctions have the same critical current $I_c$ and junction capacitance $C$, and thus the same Josephson energy $E_J \equiv I_c \Phi_0 / 2\pi$ and charging energy $E_C \equiv e^2/2C$. The third junction is smaller in area by a factor $\alpha$, resulting in a reduced critical current $\alpha I_c$ and reduced junction capacitance $\alpha C$. The corresponding energy scales are $\alpha E_J$ and $E_C/\alpha$.

The flux quantization condition gives $\varphi_1 - \varphi_2 + \varphi_3 = 2\pi f_e$, where $f_e \equiv \Phi_e/\Phi_0$ is the external magnetic flux threading the loop in units of the superconducting flux quantum $\Phi_0 = h/2e$. When $\alpha > 0.5$ and $f_e \approx 0.5$, the potential energy of the qubit assumes a double-well profile. The wells are associated with clockwise and counterclockwise circulating currents tunable by the applied magnetic flux. These diabatic circulating-current states tunnel-couple with a strength depending on the height of the inter-well barrier or the scale factor $\alpha$.

Tuning $f_e$ tilts the double well potential. In the vicinity of $f_e \approx 0.5$, the circulating currents are of opposite sign and essentially equal magnitude. Consequently, in this limit, higher energy levels play little role and the two-level approximation is a good approximation to the full Hamiltonian. However, even for the conventional flux qubit, higher energy levels become important for flux biases far from $f_e \approx 0.5$, where the double-well potential is tilted to such a degree that its higher levels influence the circulating currents [4,5]. The region about $f_e \approx 0.5$ for which the two-level system is a good approximation is reduced as $\alpha$ decreases.

## Supplementary Note 4.2  C-shunt flux qubit: quasistatics

The C-shunt flux qubit has three features that are distinctly different from the conventional design: (i) lower $I_c$, (ii) lower $\alpha$, typically $\alpha < 0.5$ (defined for three-junction qubits) and (iii) an additional large capacitor $C_{sh} = \zeta C$ ($\zeta \gg 1$) shunting the smaller junction. As we show below, both (i) and (ii) lead to reduced sensitivity to flux noise, while (iii) reduces sensitivity to charge noise.

Following a recipe similar to that presented in Ref. [3], the three-junction capacitively-shunted flux qubit can be described by a Hamiltonian consisting of the kinetic and potential part:

$$\hat{\mathcal{H}} = \hat{T} + \hat{U} ,$$
$$\hat{T} = \frac{1}{2}(\hat{\mathbf{Q}} + \hat{\mathbf{q}})^{\mathrm{T}} \mathbf{C}^{-1} (\hat{\mathbf{Q}} + \hat{\mathbf{q}}) ,$$
$$\hat{U} = E_{\mathrm{J}} \{ 2 + \alpha - \cos\varphi_1 - \cos\varphi_2 - \alpha \cos(2\pi f_e + \varphi_1 - \varphi_2) \} . \tag{4}$$

In the kinetic part, $\hat{\mathbf{Q}}$ and $\hat{\mathbf{q}}$ are the charges and induced charges on the islands:

$$\hat{\mathbf{Q}} = -\mathrm{i}\, 2e \begin{pmatrix} \frac{\partial}{\partial \varphi_1} \\ \frac{\partial}{\partial \varphi_2} \end{pmatrix} \quad \text{and} \quad \hat{\mathbf{q}} = \begin{pmatrix} q_{\mathrm{A}} \\ q_{\mathrm{B}} \end{pmatrix} ; \quad \mathbf{C} = C \begin{pmatrix} \zeta + 1 + \alpha & -(\zeta + \alpha) \\ -(\zeta + \alpha) & \zeta + 1 + \alpha \end{pmatrix} . \tag{5}$$

The kinetic energy represents the total electrostatic energy stored in the capacitors, and is dominated by the shunt capacitor, since $\zeta \gg 1, \alpha$. The shunt capacitor largely reduces the effective charging energy, causing the system less sensitive to charge fluctuations. Compared with the circuit model in Ref. [3], there is no explicit gate electrode in our circuit. The induced charge, however, is included for modelling the charge noise, and will be discussed later in this section. In the static case, $q_{\mathrm{A}} = q_{\mathrm{B}} = 0$. The potential energy sums up the Josephson energy stored in the junctions. It assumes a two-dimensional periodic profile. When $\alpha \leq 0.5$, as in this C-shunt design, there is only one well in each unit cell (see Supplementary Figures 4a and 4b), a distinction from the double-well profile in the conventional case. This leads to a smaller circulating current for the C-shunt flux qubit, reducing its sensitivity to flux fluctuations.

By choosing $\varphi_{\mathrm{p}} = (\varphi_1 + \varphi_2)/2$ and $\varphi_{\mathrm{m}} = (\varphi_1 - \varphi_2)/2$ as coordinates and using the Cooper-pair number operators $\hat{n}_\sigma = -\mathrm{i}\,\partial/\partial\varphi_\sigma$ ($\sigma = \mathrm{p}, \mathrm{m}$), we have the reduced Hamiltonian

$$\hat{\mathcal{H}} = \frac{1}{2} E_{\mathrm{C,p}} \hat{n}_{\mathrm{p}}^2 + \frac{1}{2} E_{\mathrm{C,m}} \hat{n}_{\mathrm{m}}^2 + E_{\mathrm{J}} \{ 2 + \alpha - 2\cos\varphi_{\mathrm{p}} \cos\varphi_{\mathrm{m}} - \alpha \cos(2\pi f_e + 2\varphi_{\mathrm{m}}) \} , \tag{6}$$

where $E_{\mathrm{C,p}} = e^2/(C/2)$ and $E_{\mathrm{C,m}} = e^2/C(\zeta + \alpha + 1/2)$ are the effective charging energy for the p-mode and m-mode respectively. Ideally, because the introduction of the shunt capacitor ensures that $E_{\mathrm{C,p}} \approx 2\zeta E_{\mathrm{C,m}} \gg E_{\mathrm{C,m}}$, we can safely omit the p-mode, since the characteristic frequency $\Omega_{\mathrm{p}} = \sqrt{E_{\mathrm{C,p}} E_{\mathrm{J,p}}}/\hbar \approx 2\pi \times 50\,\mathrm{GHz}$ is much higher. Here, $E_{\mathrm{J,p}} = 2E_{\mathrm{J}}$ is the effective p-mode Josephson energy. In a realistic design, one must also assess the role of additional plasma modes, for example, due to capacitance to ground, and their characteristic frequencies. The simplified Hamiltonian becomes

$$\hat{\mathcal{H}}_{\mathrm{m}} = \frac{1}{2} E_{\mathrm{C,m}} \hat{n}_{\mathrm{m}}^2 + E_{\mathrm{J}} \{ -2\cos\varphi_{\mathrm{m}} + \alpha \cos(2\pi f_b + 2\varphi_{\mathrm{m}}) \} , \tag{7}$$

where $f_b = f_e - 0.5$ is the reduced external flux bias away from one-half flux quantum.

To understand the energy structure of the system, we Taylor-expand the potential function $U(\varphi_{\mathrm{m}}) = E_{\mathrm{J}} \{ -2\cos\varphi_{\mathrm{m}} + \alpha \cos(2\pi f_b + 2\varphi_{\mathrm{m}}) \}$ around the well minimum $\varphi_{\mathrm{m}}^*$,

$$U(\varphi_{\mathrm{m}}) = \sum_{k=0}^{\infty} U^{(k)} (\varphi_{\mathrm{m}} - \varphi_{\mathrm{m}}^*)^k , \tag{8}$$

where $U^{(k)} = (1/k!)(\partial^k U/\partial\varphi_m^k|_{\varphi_m=\varphi_m^*})$. For a given $f_b$, $\varphi_m^*$ is single-valued within the range $\varphi_m \in [-\pi, \pi]$ and obtainable by solving the equation $U'(\varphi_m^*) = 0$. By definition, $U^{(0)}$ is a constant and $U^{(1)} = 0$. Therefore, we can rewrite Eq. (7) as:

$$\hat{\mathcal{H}}_m = \hat{\mathcal{H}}_0 + \hat{V}$$
$$= \left(\frac{1}{2}E_{C,m}\hat{n}_{m'}^2 + \frac{1}{2}E_{J,m}\hat{\varphi}_{m'}^2\right) + \sum_{k=3}^{\infty} U^{(k)}\hat{\varphi}_{m'}^k$$
$$= \hbar\Omega_m^{(0)}(\hat{b}^\dagger\hat{b} + \frac{1}{2}) + \sum_{k=3}^{\infty} U^{(k)}\varphi_z^k(\hat{b} + \hat{b}^\dagger)^k \ . \tag{9}$$

On the second line, $E_{J,m} = 2U^{(2)}$ is the effective Josephson energy, $\hat{\varphi}_{m'} = \varphi_{m'} = \varphi_m - \varphi_m^*$ and $\hat{n}_{m'} = -\mathrm{i}\partial/\partial\varphi_{m'}$. We group the quadratic potential term with the kinetic part so that their combination $\mathcal{H}_0$ describes a one-dimensional harmonic oscillator. The last line expresses the Hamiltonian in terms of the raising ($\hat{b}^\dagger$) and lowering ($\hat{b}$) operators of the harmonic oscillator. The oscillator frequency is $\Omega_m^{(0)} = \sqrt{E_{C,m}E_{J,m}}/\hbar$. The quantum ground-state phase uncertainty is $\varphi_z = (E_{C,m}/4E_{J,m})^{1/4}$ [6]. After the replacement $\hat{\varphi}_{m'} = \varphi_z(\hat{b} + \hat{b}^\dagger)$, the perturbation matrix $V$ can be conveniently computed in the space spanned by the Fock states of $\mathcal{H}_0$. Note that $\hbar\Omega_m^{(0)}$, $U^{(k)}$ and $\varphi_z$ are all $f_b$-dependent. The eigenstates can be solved for a given order of expansion. In general, higher-power expansion terms introduce anharmonicity as well as modulate the harmonic frequency. For simplicity and without loss of generality in the approach, we restrict ourselves to the subsystem spanned by the first-three levels $\{|g\rangle, |e\rangle, |f\rangle\}$. Within a first-order perturbation analysis, we have the approximate Hamiltonian

$$\hat{\mathcal{H}} = \hbar\left(\Omega\hat{b}^\dagger\hat{b} + \frac{\mathcal{A}}{2}\hat{b}^\dagger\hat{b}^\dagger\hat{b}\hat{b}\right) \ , \tag{10}$$

where $\Omega$ and $\mathcal{A}$ are respectively the $f_b$-dependent harmonicity and anharmonicity. As an important example, they can be derived at $f_b = 0$, where $\varphi_m^* = 0$. The expansion terms then become

$$U^{(2)}/E_J = 1 - 2\alpha \ ,$$
$$U^{(3)}/E_J = 0 \ ,$$
$$U^{(4)}/E_J = \frac{8\alpha - 1}{12} \ ,$$
$$U^{(5)}/E_J = 0 \ ,$$
$$\ldots \tag{11}$$

All odd-power terms disappear due to symmetry of the potential at $f_b = 0$. The term $U^{(2)}$ provides a quadratic potential and thus harmonicity when $\alpha < 0.5$, which is consistent with the "slightly anharmonic oscillator" model developed here. When $\alpha > 0.5$, as in the conventional flux qubit, the potential becomes a concave function with a rising barrier which splits the landscape into double wells. The quartic potential from $U^{(4)}$ is the leading term that introduces anharmonicity. Keeping terms up to $U^{(4)}$, we have $\hbar\Omega = \sqrt{E_{C,m}E_{J,m}} + \lambda E_{C,m}$ and $\hbar\mathcal{A} = \lambda E_{C,m}$, where $E_{J,m} = 2(1-2\alpha)E_J$ and $\lambda = \frac{8\alpha - 1}{8(1-2\alpha)}$. Our present C-shunt designs are within the regime of $\alpha > 0.125$, giving a positive anharmonicity in contrast to the negative one which is characteristic in transmon qubits.

## Supplementary Note 4.3  C-shunt flux qubit: time-dependent fluctuations

We now consider flux noise and charge noise within this analytical model. Small flux fluctuations ($\delta f$) affect this anharmonic system by modulating the potential energy. That is

$$\begin{aligned}\delta\hat{\mathcal{H}}_f = \delta\hat{U} &= -2\pi\alpha E_{\rm J}\sin(2\pi f_{\rm b} + 2\varphi_{\rm m})\delta f \\ &= -2\pi\alpha E_{\rm J}\{\sin\phi\,\cos(2\varphi_{\rm m'}) + \cos\phi\,\sin(2\varphi_{\rm m'})\}\delta f \\ &\approx -2\pi\alpha E_{\rm J}\{\sin\phi\,(1-2\varphi_{\rm z}^2(\hat{b}+\hat{b}^\dagger)^2) + \cos\phi\,2\varphi_{\rm z}(\hat{b}+\hat{b}^\dagger)\}\delta f\;,\end{aligned} \quad (12)$$

where $\phi = \phi(f_{\rm b}) = 2\pi f_{\rm b} + 2\varphi_{\rm m}^*(f_{\rm b})$. The $(\hat{b}+\hat{b}^\dagger)^2$ term contains longitudinal modulation while the $(\hat{b}+\hat{b}^\dagger)$ term provides only transverse modulation, that is, a coupling matrix element between adjacent levels.

In the spectral domain, fluctuations near frequency $\Omega$ are of particular interest because of their ability to induce g−e transitions, and thus play an essential role in qubit driving and relaxation. Consider a coherent drive $\delta f \propto \cos(\omega_{\rm d} t)$ with driving frequency $\omega_{\rm d} \approx \Omega$. In the interaction frame, one can show that both the fast longitudinal oscillations and transverse counter-rotating terms are negligible in the weak driving limit. This simplifies the three-level interaction-frame Hamiltonian to

$$\tilde{\hat{\mathcal{H}}} = \hbar\begin{pmatrix} 2\Delta\omega + \mathcal{A} & \sqrt{2}\,\Omega_{\rm R}/2 & 0 \\ \sqrt{2}\,\Omega_{\rm R}/2 & \Delta\omega & \Omega_{\rm R}/2 \\ 0 & \Omega_{\rm R}/2 & 0 \end{pmatrix}\;, \quad (13)$$

where $\Delta\omega = \Omega - \omega_{\rm d}$ is the detuning of the drive from the g−e transition frequency. In the regime of weak drive, namely $\omega_{\rm q}, \mathcal{A} \gg \Omega_{\rm R}$, the system is effectively protected from transitions to the third level. Therefore, we can further reduce the system to two levels,

$$\hat{\mathcal{H}} = \frac{1}{2}\left(\hbar\omega_{\rm q}\,\hat{\sigma}_{\rm z} + I_{\rm m}\Phi_0\,\delta f\,\hat{\sigma}_{\rm x}\right)\;, \quad (14)$$

where $\omega_{\rm q} = \omega_{\rm q}(f_{\rm b}) = \Omega$ as in Eq. (10) and $I_{\rm m} = I_{\rm m}(f_{\rm b}) = -8\pi\alpha\,\varphi_{\rm z}\cos\phi\,E_{\rm J}/\Phi_0$. Here, $I_{\rm m}$ is the $f_{\rm b}$-dependent current difference between the parameterized circulating-current states (See section Supplementary Note 3). The scaling $I_{\rm m} \propto \alpha E_{\rm J}^{3/4} E_{\rm C}^{1/4} \zeta^{-1/4}$ indicates the efficiency of reducing flux-noise sensitivity by lowering $\alpha$ and $E_{\rm J}$, as we implemented in the C-shunt design. For example, qubit sample B in our device (see main text) has $\alpha \approx 0.4$, $E_{\rm J}/h \approx 65\,{\rm GHz}$ and $\varphi_{\rm z} \approx 0.28$, leading to $I_{\rm m} \approx 55\,{\rm nA}$.

On the other hand, charge fluctuations ($\delta q_{\rm A}$ and $\delta q_{\rm B}$) invoke perturbation via the kinetic energy $\hat{T}$. Expanding $\hat{T}$ in Eq. (4) and ignoring the p-mode (as before), we have the perturbation Hamiltonian,

$$\begin{aligned}\delta\hat{\mathcal{H}}_q = \delta\hat{T} &= \delta\hat{\mathbf{q}}^{\rm T}\mathbf{C}^{-1}\hat{\mathbf{Q}} \\ &= -\frac{e}{C(2\zeta+2\alpha+1)}[-{\rm i}\,n_{\rm z}(\hat{b}-\hat{b}^\dagger)](\delta q_{\rm A} - \delta q_{\rm B})\;.\end{aligned} \quad (15)$$

On the second line, we used the transformation $\hat{n} = -{\rm i}\,n_{\rm z}(\hat{b}-\hat{b}^\dagger)$, where $n_{\rm z} = (E_{\rm J,m}/4E_{\rm C,m})^{1/4}$ is the quantum ground-state uncertainty in Cooper-pair number. Similar to the flux-noise argument above, charge fluctuations also connect to the system transversely, and moreover, are also orthogonal to the flux. In addition, the perturbation depends only on the differential mode of the induced charges between islands A and B, the branch charge across the small junction.

Adding $\delta\hat{\mathcal{H}}_q$ to the two-level approximated Hamiltonian in Eq. (14), we have

$$\hat{\mathcal{H}} = \frac{1}{2}\left(\hbar\omega_{\rm q}\,\hat{\sigma}_{\rm z} + I_{\rm m}\Phi_0\,\delta f\,\hat{\sigma}_{\rm x} + n_{\rm z}E_{\rm C,m}\delta n_{\rm m}\,\hat{\sigma}_{\rm y}\right)\;, \quad (16)$$

where $\delta n_{\rm m} = (\delta q_{\rm A} - \delta q_{\rm B})/(-e)$ is the differential electron number fluctuation. The charge-noise sensitivity is $n_{\rm z}E_{\rm C,m} \propto E_{\rm C,m}^{3/4} \propto \zeta^{-3/4}$. Therefore, the introduction of a large shunt capacitor makes the system less sensitive to charge fluctuations. In qubit sample B (see main text), $n_z E_{\text{C,m}}/h \approx 1.2\,\text{GHz}$. Assuming Johnson-Nyquist noise from a 377-$\Omega$ resistor (free-space impedance) and a parameterized gate capacitance of $0.03\,C = 0.12\,\text{fF}$, we find a charge-noise-limited $T_1$ of about $60\,\mu$s.

The capacitive coupling of the qubit loop to the resonator can be modelled in a similar way as the node charges. The voltage fluctuations on the coupling capacitor $C_{\text{g}}$ induce charge fluctuations, and hence enter the Hamiltonian through the same channel as charge. Consequently, the interaction Hamiltonian between the qubit and resonator can be written as

$$\hat{\mathcal{H}}_{\text{q-r}} = \hbar g\, \hat{\sigma}_{\text{y}}\, (\hat{a}^\dagger + \hat{a})\,, \tag{17}$$

where $g$ is the coupling strength.

## Supplementary Note 5    Simulation of the Full System Hamiltonian

Supplementary figure 5a shows a schematic of the configuration for the devices described in this work, where the five islands labelled numerically in white text (1-5) follow the electrostatic description of Ref. [1] with island 4 the center conductor of the coplanar waveguide resonator. Supplementary figure 5b shows an equivalent circuit in which islands 1 and 5 have been grounded (equivalent to neglecting the parasitic stripline modes of the resonator). Filled black circles in this circuit indicate the five canonical node variables that would be required for its full quantum description. Junction capacitances are labelled $C_{\text{Ja}}$ and $C_{\text{Jb}}$ for the small and large junctions, respectively, and $L_{\text{g}}$ is the geometric inductance of the qubit loop. Supplementary figures 5c and 5d show successive (approximate) reductions of the circuit to four and three node variables (labelled in blue), respectively. Bold lines in these schematics indicate the chosen spanning tree (equivalent to a choice of gauge) [7, 8].

The quantities in panels (b) and (c) are related to those in (b) according to

$$\begin{aligned}
C_{\text{T}} &= C_{13} + C_{35} + \frac{C_{34}}{C_4^{\text{tot}}}(C_{14} + C_{15}) \\
C_{\text{a}} &= C_{23} + \frac{C_{24}}{C_4^{\text{tot}}}C_{34} \\
C_{\text{b}} &= C_{12} + C_{25} + \frac{C_{24}}{C_4^{\text{tot}}}(C_{14} + C_{15}) \\
C_{\text{sh}} &= C_{\text{A}} + \frac{C_{\text{T}}C_{\text{B}}}{C_{\text{T}} + C_{\text{B}}}
\end{aligned} \tag{18}$$

where $C_4^{\text{tot}} \equiv C_{14} + C_{24} + C_{34} + C_{45}$. The dimensionless capacitive division factor by which the resonator voltage is coupled to the qubit is given by:

$$\begin{aligned}
\beta_{\text{c}} =&\ [2C_{34}(C_{12} + C_{25}) - 2C_{24}(C_{13} + C_{35})] \\
\times&\ \big\{2C_{24}C_{34} + (C_{24} + C_{34})[2(C_{23} + C_{\text{Ja}}) + C_{\text{Jb}}] + (C_{13} + C_{35})[2(C_{23} + C_{24} + C_{\text{Ja}}) + C_{\text{Jb}}] \\
&+ (C_{12} + C_{25})[2(C_{13} + C_{23} + C_{34} + C_{35} + C_{\text{Ja}}) + C_{\text{Jb}}]\big\}^{-1}.
\end{aligned} \tag{19}$$

Note that the junction capacitances influence this value, in particular when the shunt capacitors are smaller.

Panel (c) is the minimal circuit that fully captures the experimental qubit devices (excluding the resonator), in particular because in nearly all devices the total $C_{\text{sh}}$ is determined by the combination of $C_{\text{T}}$, $C_{\text{B}}$, and $C_{\text{a}}$. The approximation inherent in (d) arises from the fact that the two fundamental Josephson-like modes of the circuit (the qubit mode in which nodes 1 and 2 oscillate out of phase, and the "plasma" mode in which they oscillate in phase) couple to the same shunt capacitance $C_\text{sh}$. By contrast, in the circuit of panel (c), the plasma mode may couple to a larger capacitance $C_\text{B} + C_\text{T} > C_\text{a}$ than the qubit mode. The result is that, unlike (c) where increasing the shunt capacitance leaves the plasma mode frequency far above the region of interest, in the more realistic circuit of (d) the plasma mode frequency shifts down as the total shunt capacitance is increased, due to the effect of $C_\text{B}$ and $C_\text{T}$. For the purpose of the present work, however, it turns out that the circuit of panel (d) gives a good approximation (5% or better) to all quantities considered, except inelastic quasiparticle tunneling matrix elements, where a correct treatment of tunneling events through the two larger junctions precludes grounding one side of those junctions. Note that it is necessary to include the geometric loop inductance (though its effect on the energy levels is small) to account for the flux noise matrix elements to this accuracy.

The Hamiltonians for the two circuits are

$$\begin{aligned}
\hat{\mathcal{H}}_3 &= 4\hat{\mathbf{n}} \cdot \mathbf{E}_\text{C3} \cdot \hat{\mathbf{n}} + E_{\text{J}\alpha}\big[1 - \cos(\hat{\varphi}_2 - \hat{\varphi}_1 + \varphi_\text{e})\big] \\
&\quad + E_{\text{Jb}}\big[2 - \cos(\hat{\varphi}_2 - \hat{\varphi}_3) - \cos(\hat{\varphi}_1)\big]
\end{aligned} \quad (20)$$

$$\begin{aligned}
\hat{\mathcal{H}}_4 &= 4\hat{\mathbf{n}} \cdot \mathbf{E}_\text{C4} \cdot \hat{\mathbf{n}} + E_{\text{J}\alpha}\big[1 - \cos(\hat{\varphi}_2 - \hat{\varphi}_1 + \varphi_\text{e})\big] \\
&\quad + E_{\text{Jb}}\big[2 - \cos(\hat{\varphi}_2 - \hat{\varphi}_3) - \cos(\hat{\varphi}_4 - \hat{\varphi}_1)\big] + \frac{E_\text{L}}{2}(\hat{\varphi}_3 - \hat{\varphi}_4)^2
\end{aligned} \quad (21)$$

where $\hat{\mathbf{n}} \equiv \{\hat{n}_i\}$ is the vector node charge operator, $\mathbf{E}_\text{C3,4} \equiv e^2/2 \cdot \mathbf{C}_{3,4}^{-1}$ is the inverse charging energy matrix, $E_{\text{J}\alpha}$ and $E_{\text{Jb}}$ are the Josephson energies of the small and large junctions, respectively, $\hat{\varphi}_i$ are the node phase operators satifying $[\hat{\varphi}_j, \hat{n}_k] = i\delta_{jk}$, and $\varphi_\text{e} \equiv 2\pi\Phi_\text{e}/\Phi_0$ is the dimensionless external flux through the qubit loop. Finally, $E_\text{L} \equiv (\Phi_0/2\pi)^2/L_\text{g}$ is the characteristic inductive energy scale for the geometric loop inductance $L_\text{g}$. The capacitance matrices for the two circuits are given by:

$$\mathbf{C}_3 = \begin{pmatrix} C_\text{Jb} + C_\text{Ja} + C_\text{sh} & -C_\text{Ja} - C_\text{sh} & 0 \\ -C_\text{Ja} - C_\text{sh} & C_\text{Jb} + C_\text{Ja} + C_\text{sh} & -C_\text{Jb} \\ 0 & -C_\text{Jb} & C_\text{Jb} \end{pmatrix} \quad (22)$$

$$\mathbf{C}_4 = \begin{pmatrix} C_\text{B} + C_\text{Jb} + C_\text{Ja} + C_\text{a} & -C_\text{Ja} - C_\text{a} & 0 & -C_\text{Jb} \\ -C_\text{Ja} - C_\text{a} & C_\text{T} + C_\text{Jb} + C_\text{Ja} + C_\text{a} & -C_\text{Jb} & 0 \\ 0 & -C_\text{Jb} & C_\text{Jb} & 0 \\ -C_\text{Jb} & 0 & 0 & C_\text{Jb} \end{pmatrix} \quad (23)$$

To diagonalize the full Hamiltonians for the two circuits, we first set the phase across the inductor to zero and diagonalize the resulting Hamiltonian (which has either two or three node variables) in a truncated charge basis containing the states: $-10, -9, ..., 1, 0, 1, ..., 9, 10$ for each island (in units of Cooper pairs), having dimension $21^2 = 441$ or $21^3 = 9261$. We use the resulting set of eigenstates $\Psi_m^{(0)}$ and eigenenergies $E_m^{(0)}$ to re-express the full Hamiltonian in a product state basis: $\Psi_m^{(0)} \otimes |\nu\rangle$ where $|\nu\rangle$ are linear oscillator states resulting from the loop inductance $L_\text{g}$ and the shunt capacitance across it. The oscillator basis is truncated at $\nu \leq 3$, and the $L_\text{g} = 0$ qubit basis is truncated at $m \leq 75$ for the three-node circuit, and $m \leq 300$ for the four node circuit. The increased basis size for the latter case is necessary because the four-node circuit does not have an inductance for its common mode (the one in which all islands oscillate together relative to ground), and therefore no potential energy, resulting in simple, charge eigenstates. The resulting Hilbert space dimensions for the two circuits are $3 \times 75 = 225$ and $3 \times 300 = 900$, much smaller than that which would have otherwise been required: $21^2 \times 3 = 1323$ and $21^3 \times 3 = 27783$ states. This method is a useful way to efficiently include linear inductances into Josephson quantum circuits [9].

Once the eigenenergies and eigenstates of the qubit circuits are determined, this system is again truncated (typically at ~10 qubit energy levels) and then coupled to the resonator using the Hamiltonian

$$\hat{\mathcal{H}}_\text{q-r} = 2e\,(\hat{n}_1 - \hat{n}_2) \cdot \beta_\text{c} V_\text{r0}(\hat{a}^\dagger + \hat{a}), \quad (24)$$

where $\beta_c$ is obtained from Eq. (19) above, and $V_{r0} = \omega_r \sqrt{2\hbar Z_r}$ is the rms ground state resonator mode voltage for a mode impedance $Z_r$. A truncated basis of up to $\sim 5$ photons in the resonator mode is then used to diagonalize the resulting dressed Hamiltonian.

## Supplementary Note 6     High-Power Spectrum and Higher-Level Qubit Transitions

To characterize higher levels of the qubit, we perform high-power qubit spectroscopy (Supplementary Figure 6). The measurement consists of scanning the drive frequency ($\omega_d$) as a function of the flux bias with a much higher power than that used in the main text in Fig. 2a. Strong driving reveals more qubit transitions in the spectrum, including the first four qubit levels, single and multi-photon transitions, and resonator-mediated transitions. The same simulation used to develop our noise models is used to add the solid lines in Supplementary Figure 6. To match the measured transition frequencies optimally across the entire flux bias range, we generally need to tune somewhat the design parameters (Supplementary Table 1), typically by $10-25\%$. Once adjusted, the fitting is very good and reproduces the measured spectroscopy over a wide range of flux and frequency values.

To characterize the anharmonicity accurately, we use a two-tone low-power pulse technique to measure 0–1 and 1–2 transitions sequentially (Supplementary Figure 7). A short $\pi$-pulse at 0–1 transition frequency first prepares the qubit at $|1\rangle$ state. Then, a low-power frequency scan resolves the 0–1 and 1–2 transitions. The frequency difference gives the anharmonicity which is 490 MHz for Qubit B. This is somewhat less than the 830 MHz predicted solely from the design values (Supplementary Table 1). Although the measured value is off by approximately 330 MHz from the anharmonicity predicted directly from the design values, it is important to keep in mind that this few-hundred MHz difference is on top of qubit frequencies more than 10 times larger (i.e., 4.7-5.2 GHz) and, in this sense, the error is rather small.

## Supplementary Note 7     Qubit Parameters

The parameters for the 22 qubits studied in the work are listed in Supplementary Table 1.

## Supplementary Note 8     Noise Models

The total $T_1$ of the qubit is taken to be:

$$\frac{1}{T_1} = \frac{1}{T_1^\Phi} + \frac{1}{T_1^Q} + \frac{1}{T_1^{qp}} + \frac{1}{T_1^P} \tag{25}$$

where the terms come from electric noise, magnetic noise, inelastic quasiparticle fluctuations, and the Purcell effect, respectively.

The coupling of electric and magnetic noise to the qubit is modeled as shown in Supplementary Figure 8, in terms of voltage sources $\delta V_i$ weakly coupled to the circuit islands, and a current source $\delta I_l$ weakly coupled to the circuit loop. We assume the weak-coupling limit for both of these, where $C_{ci}$ are negligible compared to the corresponding node capacitances (diagonal elements of $\mathbf{C}$), and $M_c$ is negligible compared to the qubit's loop inductance $L_\text{g}$. In this limit, the electric noise can be expressed as well-defined charge fluctuations $\delta Q_i$ for node $i$ (with magnitudes independent of the node capacitances), and the magnetic noise as flux fluctuation through the loop $\delta \Phi_l$ (with magnitude independent of the loop inductance). The coupling Hamiltonians for these fluctuations to the qubit can be written:

$$\hat{\mathcal{H}}_{\delta\Phi} = \hat{I}_l \delta\Phi_l \tag{26}$$

$$\hat{\mathcal{H}}_{\delta Q_i} = \hat{V}_i \delta Q_i \tag{27}$$

where $\hat{I}_l \equiv \Phi_0(\hat{\varphi}_3 - \hat{\varphi}_4)/2\pi/L_\text{g}$ is the loop current operator and $\hat{V}_i \equiv (\hat{\mathbf{Q}} \cdot \mathbf{C}^{-1})_i$ is the voltage operator for node $i$. The resulting contributions to the total decay rate are:

$$\frac{1}{T_1^\Phi} = \frac{2}{\hbar^2} |\langle \text{e}|\hat{I}_l|\text{g}\rangle|^2 S_\Phi(\omega_\text{q}) \tag{28}$$

$$\frac{1}{T_1^Q} = \frac{2}{\hbar^2} |\langle \text{e}|\hat{\mathbf{V}}|\text{g}\rangle|^2 S_Q(\omega_\text{q}) \tag{29}$$

For the Purcell-enhanced decay rate of the qubit excited state, we use the expression:

$$\frac{1}{T_1^\text{P}} = |\langle \tilde{\text{e}}_{1\gamma}|(\hat{a} + \hat{a}^\dagger)|\tilde{\text{g}}_{1\gamma}\rangle|^2 \kappa \tag{30}$$

where $\kappa$ is the resonator decay rate and $|\tilde{\text{g}}_{1\gamma}\rangle, |\tilde{\text{e}}_{1\gamma}\rangle$ are the two dressed energy eigenstates of the qubit/resonator system in the one-photon subspace.

As shown in purple in Supplementary Figure 8, the quasiparticle noise contribution to the decay rate is modeled, following Ref. 10, as a parallel admittance $Y_\text{qp}$ and corresponding current fluctuations $\delta I$. This approximate description is justified since $Y_\text{qp}(\omega_\text{q}) \ll Y_\text{J}(\omega_\text{q})$ where $Y_\text{J}$ is the junction impedance. The resulting decay rate is [10, 11]:

$$\frac{1}{T_1^\text{qp}} = \frac{1}{4} \sum_{k=1}^{3} |\langle \tilde{\text{e}}_{1k}|\hat{t}_\text{qp}^k e^{i\varphi_k/2} - \hat{t}_\text{qp}^{k\dagger} e^{-i\varphi_k/2}|\tilde{\text{g}}_{1k}\rangle|^2 S_\text{qp}^k(\omega_\text{q}) \tag{31}$$

where the sum is over the three junctions, the operator $\hat{t}_\text{qp}^k$ transfers a quasiparticle through junction $k$, the phase offsets are $\varphi_k = \varphi_\text{e}$ for the small junction and zero otherwise, and $|\tilde{g}_{1k}\rangle, |\tilde{e}_{1k}\rangle$ are qubit energy eigenstates in the presence of a single quasiparticle on one electrode of junction $k$. The effective quasiparticle current noise spectral density $S_\text{qp}^k$ for junction $k$ is given approximately by [10, 11]:

$$S_\text{qp}^k = x_\text{qp} \frac{E_{\text{J}k}}{h} \sqrt{\frac{8\Delta_\text{Al}}{\omega_\text{q}}} \tag{32}$$

where $x_\text{qp}$ is the dimensionless quasiparticle density (scaled by the density of superconducting electrons), and $\Delta_\text{Al}$ is the superconducting energy gap of aluminum. The eigenstates of the qubit in the presence of a single quasiparticle on each circuit node are obtained using a modified charge representation for that node with 20 basis states: $-19/2, -17/2, ..., -1/2, 1/2, ..., 17/2, 19/2$ (in units of Cooper pairs). The Hamiltonian is separately diagonalized for each single-quasiparticle configuration using the methods described above. Then the matrix elements in Eq. (31) above are evaluated between eigenstates of these different configurations, corresponding to a single quasiparticle moving through each of the three Josephson junctions in the loop. The resulting decay rates associated with inelastic tunneling through each junction are summed to produce the total rate. Note that we neglect processes associated with the presence of two quasiparticles simultaneously, which is justified based on our observation of $\bar{n}_\text{qp} \approx 0.26$.

## Supplementary Note 9   Definitions of Power Spectral Densities

The spectral data shown in Fig. 3b in the main text represent a symmetrized PSD, the Fourier transform of the symmetrized autocorrelation function,

$$S_\lambda(\omega) = \int_{-\infty}^{\infty} d\tau \, \exp(-i\omega\tau) \frac{1}{2} \langle \hat{\lambda}(0)\hat{\lambda}(\tau) + \hat{\lambda}(\tau)\hat{\lambda}(0) \rangle \,, \tag{33}$$

where $\hat{\lambda}$ is an operator representing the parameter (e.g., flux, charge, current, voltage, ...) that is fluctuating. Clearly, by definition, $S_\lambda(-\omega) = S_\lambda(\omega)$, and the effects of noise in the classical ($\hbar\omega_\mathrm{q} \ll k_\mathrm{B} T$) and quantum ($\hbar\omega_\mathrm{q} \gg k_\mathrm{B} T$) regimes are represented equivalently at both positive and negative frequencies.

In comparison, the unsymmetrized definition for the PSD,

$$S_\lambda^\mathrm{U}(\omega) = \int_{-\infty}^{\infty} d\tau \, \exp(-i\omega\tau) \langle \hat{\lambda}(0)\hat{\lambda}(\tau) \rangle \,, \tag{34}$$

explicitly distinguishes energy absorption and emission by the qubit: noise at negative frequencies corresponds to energy absorption by the qubit from its environment, whereas noise at positive frequencies corresponds to energy emitted by the qubit to its environment.

By definition, $S_\lambda(\omega) = \frac{1}{2}\big(S_\lambda^\mathrm{U}(\omega) + S_\lambda^\mathrm{U}(-\omega)\big)$, and so either can be used to describe the data presented in this work. Here, we elected to use the symmetrized PSD $S_\lambda(\omega)$, because it has the useful attribute that it can directly connect noise data measured in the classical and quantum noise regimes (see Figure 3b in the main text).

The emission and absorption rates for a system with level-spitting frequency $\omega_\mathrm{q}$ are related to the positive and negative part of the spectrum respectively,

$$\Gamma_- = \sum_\lambda \frac{1}{4} k_\lambda^2 \, S_\lambda^\mathrm{U}(\omega_\mathrm{q}) \,,$$

$$\Gamma_+ = \sum_\lambda \frac{1}{4} k_\lambda^2 \, S_\lambda^\mathrm{U}(-\omega_\mathrm{q}) \,. \tag{35}$$

Here, $\Gamma_{-(+)}$ is the energy emission (absorption) rate to (from) the environment, and $k_\lambda$ is the system noise sensitivity, defined as the derivative of the transversal energy change in units of angular frequency with respect to $\lambda$, i.e., $k_\lambda = \frac{1}{\hbar} \frac{\partial |\mathcal{H}_\perp|}{\partial \lambda}$. It is related to the transition dipole matrix element – as presented in Eq. (2) in the main text – by $k_\lambda = \frac{2}{\hbar} |\langle e| \hat{D}_\lambda |g\rangle|$. The factor 2 can be understood intuitively as the dipole matrix element addressing only one of the two transverse matrix elements, whereas the total transverse energy sensitivity to noise is related to both off-diagonal elements. Consider, for example, the traditional flux qubit with the two-level-system Hamiltonian $\hat{\mathcal{H}} = \hbar\Delta\,\hat{\sigma}_\mathrm{z}/2 + \hbar\varepsilon\,\hat{\sigma}_\mathrm{x}/2$, where $\hbar\varepsilon = 2I_\mathrm{p}\Phi_\mathrm{b}$ is the energy bias, $I_\mathrm{p}$ is the persistent current and $\Phi_\mathrm{b}$ is the flux bias [12]. While the current operator matrix element $|\langle e|\hat{I}_\lambda|g\rangle| \approx I_\mathrm{p}$, the flux-noise sensitivity $k_\Phi = |\frac{\partial \varepsilon}{\partial \Phi}| = 2I_\mathrm{p}/\hbar$, giving the factor 2. $k_\lambda$ is convenient in practice, as it stands for the change in energy per unit change in $\lambda$. For example, $k_\Phi = 2\pi \times 1000\,\mathrm{GHz}/\Phi_0$ means $\varepsilon/2\pi = 1\,\mathrm{MHz}$ for a $1\,\mu\Phi_0$ change in flux.

The decay rate we observe in the inversion-recovery experiment (Fig. 2(c) in the main text) corresponds to the sum of emission and absorption rates,

$$\begin{aligned}\frac{1}{T_1} = \Gamma_1 &= \Gamma_- + \Gamma_+ \\ &= \sum_\lambda \frac{1}{4}k_\lambda^2 \left(S_\lambda^{\rm U}(\omega_{\rm q}) + S_\lambda^{\rm U}(-\omega_{\rm q})\right) \\ &= \sum_\lambda \frac{|\langle e|\hat{D}_\lambda|g\rangle|^2}{\hbar^2} \left(S_\lambda^{\rm U}(\omega_{\rm q}) + S_\lambda^{\rm U}(-\omega_{\rm q})\right) \\ &= \sum_\lambda \frac{|\langle e|\hat{D}_\lambda|g\rangle|^2}{\hbar^2} \left(2S_\lambda(\omega_{\rm q})\right).\end{aligned} \qquad (36)$$

This recovers the expression in Eq. (2) in the main text. Note that the relation in Eq. (36) is frequency-independent, because it does not differentiate transition direction. To disentangle the up and down rates, one needs the information of the equilibrium population or polarization.

At equilibrium temperature,

$$\frac{S^{\rm U}(-\omega_{\rm q})}{S^{\rm U}(\omega_{\rm q})} = \exp\left(-\frac{\hbar\omega_{\rm q}}{k_{\rm B}T}\right). \qquad (37)$$

This equation indicates that, in the classical, low-frequency limit ($\hbar\omega \ll k_{\rm B}T$), $S^{\rm U}(\omega) = S^{\rm U}(-\omega) = S(\omega)$. In the quantum, high-frequency limit ($\hbar\omega \gg k_{\rm B}T$), $S^{\rm U}(\omega) \gg S^{\rm U}(-\omega)$, and so $S^{\rm U}(\omega) \to 2S(\omega)$. Therefore, a factor 2 difference arises at high frequencies between the two PSD definitions.

Our preference in using symmetrized PSD is due to the fact that $S_\lambda(\omega_{\rm q})$ is uniformly related to both experimentally measured dephasing (low-frequency regime) and energy decay $T_1$ (high-frequency regime). The $1/f$ noise at low frequencies is classical, arising from an ensemble of fluctuators [13]. When extending this $1/f$ trend out to higher frequencies, one does not need to scale by a factor 2 to make the connection. In contrast, when using $S^{\rm U}(\omega)$, a factor 2 is required to make this connection. Intuitively, the origin of the factor 2 arises from the definition of $S^{\rm U}(\omega)$, by which the classical noise power is symmetrically distributed over both positive and negative frequencies, whereas the the quantum noise power is entirely captured at positive frequencies.

Of course, either PSD definition can be made to work, and one must simply be cognizant of the differences between the PSD definitions [14] in order to plot the data appropriately.

## Supplementary Note 10     Noise Spectroscopy via Spin Locking

### Supplementary Note 10.1     Spin locking technique

Spin-locking or $T_{1\rho}$ noise spectroscopy is an accurate method developed for resolving noise power spectral densities (PSD) by measuring qubit relaxation rates in the rotating frame during driven evolution. The spectroscopy spans the intermediate frequency range, *i.e.,* achievable Rabi frequencies, without substantially undermining the locking condition. Details of this method are discussed in Ref. [15].

When a two-level system (TLS) is driven by a weak ($\omega_{\rm R} \ll \omega_{\rm q}$) and resonant ($\omega_{\rm R} \gg \Delta\omega$) tone, evolution can be conveniently described in the rotating frame, which revolves around the z-axis at the drive frequency (Supplementary Figure 9a). It can be viewed as a fictitious TLS with a quantizing field pointing to X. The level splitting is now the (locking) Rabi frequency, $\omega_{\rm R}$, rather than $\omega_{\rm q}$ for the free-evolution case. Note that the corresponding longitudinal relaxation time, $T_{1\rho}$, is defined with respect to the new quantization axis. The source of the relaxation is noise at the Rabi frequency, transverse to the X-axis. Given the definition in Ref. [15], the rate $\Gamma_{1\rho}$ can be expressed as

$$\begin{aligned}
\Gamma_{1\rho} = \frac{1}{T_{1\rho}} &= \frac{1}{2} S_{\perp X}(\omega_R) \\
&= \frac{1}{2} \big[ S_Y(\omega_R) + S_Z(\omega_R) \big] \\
&= \frac{1}{2} \big[ \frac{1}{4} S_x(\omega_q + \omega_R) + \frac{1}{4} S_x(\omega_q - \omega_R) \big] + \frac{1}{2} S_z(\omega_R) \\
&\approx \frac{1}{4} S_x(\omega_q) + \frac{1}{2} S_z(\omega_R) \\
&= \frac{1}{2} \Gamma_1 + \Gamma_\nu \ .
\end{aligned} \quad (38)$$

Here, $\Gamma_1 = 1/T_1 = \frac{1}{2} S_x(\omega_q)$ is the qubit-frame longitudinal relaxation rate and $\Gamma_\nu = \frac{1}{2} S_Z(\omega_R)$ is the rate associated with the Rabi-frequency noise. In Eq. (38), the contributing rotating-frame noise in the second line is transformed to qubit-frame noise in the third line. The "1/4"-factor arises from the halved noise amplitude at the positive sideband.

Equation (38) suggests that the noise PSD at the Rabi frequency can be extracted by measuring the qubit-frame and rotating-frame longitudinal relaxation rates and making the appropriate subtraction. The spin-locking technique [16–18] is a straightforward way to measure the rotating-frame longitudinal relaxation. We use the standard three-pulse spin-locking sequence (Supplementary Figure 9b), under which the qubit undergoes $T_{1\rho}$ relaxation during the continuous driving pulse (Supplementary Figure 9c). The recorded decay is fit to an exponential function to derive the damping rate $\Gamma_{1\rho}$.

### Supplementary Note 10.2    Spectral density for thermal photons in a resonator

Thermal photons in a resonator with decay rate $\kappa$ have an exponential two-time photon-number autocorrelation function $C(\tau)$ in the small $\bar{n}$ limit [19, 20],

$$C(\tau) = \bar{n} \exp(-\kappa \tau). \quad (39)$$

Correspondingly, by the Weiner-Khinchin theorem, the associated power spectral density (PSD) of the thermal photons in the resonator is the fourier transform of Eq. (39),

$$S_{nn}(\omega) = \frac{2\kappa \bar{n}}{\omega^2 + \kappa^2}. \quad (40)$$

From Bloch-Redfield theory, longitudinal relaxation is connected with the two-time correlation of the transverse noise. This also applies to the rotating-frame analogue in the $T_{1\rho}$ process. The measured $T_{1\rho}$ noise spectrum is thus related to the traditional PSD. The effect from the non-Gaussian statistics in the photon noise can be ignored in such relaxation process. To derive the effective PSD as seen by the qubit, one must account for the dispersive coupling $\chi$ of the qubit to the resonator and its associated Stark shift (i.e., the frequency shift per photon), and the factor $\eta = \kappa^2/(\kappa^2 + 4\chi^2)$ that scales the effective photon population seen by the qubit. The resulting PSD is

$$S_z(\omega) = (2\chi)^2 \left[ S_{nn}(\omega) \right]_{\bar{n} \to \eta \bar{n}} = (2\chi)^2 \frac{2\kappa \eta \bar{n}}{\omega^2 + \kappa^2}. \quad (41)$$

## Supplementary Note 11  Comparing Models for Ohmic Charge Noise and Ohmic Flux Noise

In the main text Fig. 3a (device C), it is clear that our data clearly support the position that $1/f$ flux noise is a $T_1$ mechanism. However, above 3 GHz in this device, there is an ambiguity between ohmic charge noise (magenta dahsed line, Fig. 3a) and ohmic flux noise (grey dashed line, Fig. 3a). We used ohmic charge noise in our models in the main text, because there is a known physical basis for its role in relaxation from prior work and it gave a slightly better match to experimental results across all 22 devices. In Supplementary Figures 10a - 10f, we apply our model to the flux dependence of devices B and C, and to the prediction of $T_1$ at the flux insensitive points of all 22 qubits under two conditions:

- using ohmic charge noise as was done in the main text,
- using ohmic flux noise in place of ohmic charge noise.

We note that the two models give a similarly reasonable match to the data for all qubits. The one possible exception is for device C (Supplementary Figures 10a and 10b) above 6 GHz, where this device is highly sensitive to charge noise. In this region, one might make a plausible distinction between the efficacy of the two models. Making a stronger distinction between ohmic charge noise and ohmic flux noise will be a topic of future work.

## Supplementary Note 12  Thermal Photon Noise in the Low-Number Regime

The long-time behavior in Stark shift ($\Delta_{\text{Stark}}^{\text{th}}$) and dephasing rate ($\Gamma_\varphi^{\text{th}}$) due to thermal photons has a nonlinear dependence on the photon number ($\bar{n}$) in general. Equations (43)-(44) in Ref. [21] give the dependence for an arbitrary ratio between $\chi$ and $\kappa$,

$$\Delta_{\text{Stark}}^{\text{th}} = \frac{\kappa}{2} \text{Im}[\sqrt{Z}] - \chi ,$$
$$\Gamma_\varphi^{\text{th}} = \frac{\kappa}{2} \text{Re}[\sqrt{Z}] - \frac{\kappa}{2} , \tag{42}$$

where $Z = (1 + i\, 2\chi/\kappa)^2 + i\, 8\chi/\kappa$. Solving for $\sqrt{Z}$, we have

$$\text{Im}[\sqrt{Z}] = \sqrt{\frac{-(1-r^2) + \sqrt{(1+r^2)^2 + 16r^2\bar{n} + 16r^2\bar{n}^2}}{2}} ,$$
$$\text{Re}[\sqrt{Z}] = \sqrt{\frac{(1-r^2) + \sqrt{(1+r^2)^2 + 16r^2\bar{n} + 16r^2\bar{n}^2}}{2}} , \tag{43}$$

where $r = 2\chi/\kappa$.

In this work, we are focusing on the situation when $\bar{n}$ is much smaller than 1. Expanding Eq. (43) to first order in $\bar{n}$ yields:

$$\text{Im}[\sqrt{Z}] = r + \frac{2r}{1+r^2}\bar{n} ,$$
$$\text{Re}[\sqrt{Z}] = 1 + \frac{2r^2}{1+r^2}\bar{n} . \tag{44}$$

Substituting these expressions into Eqs. (42) gives Eqs. (4)-(5) in the main text. In fact, the low-number condition is $r$-dependent. The linear approximation is valid when

$$\bar{n} \ll \frac{(1+r^2)^2}{16r^2}$$
$$= \frac{1}{16}(2 + r^2 + \frac{1}{r^2}) \,. \tag{45}$$

Therefore, the condition becomes much looser when the system is in either the strong ($r>1$) or weak coupling ($r<1$) regime, meaning that the linear dependence can extend to higher photon number in both cases. The condition is tightest ($\bar{n} \ll 1/4$) when $r=1$.

## Supplementary Note 13  CPMG pulse sequence and filter functions

The Carr-Purcell [22]-Meiboom-Gill [23] (CPMG) is a dynamical decoupling pulse sequence that is the multi-pulse generalization of the Hahn spin-echo [24]. The CPMG sequence comprises equally spaced $\pi$-pulses in quadrature (with phases $90°$-shifted) with respect to the initial $\pi/2$-pulse. The technique reduces dephasing due to low-frequency noise by a coherent refocusing effect imparted by the $\pi$-pulses. The act of applying $\pi$-pulses in the time-domain can be treated in the frequency domain as a band-pass filter that shapes the noise spectra. The passband of this filter is inversely related to the spacing between adjacent $\pi$-pulses.

During free evolution, the decay function due to dephasing is written $\exp[-\xi(\tau)]$, where $\xi(\tau)$ is called the coherence function and $\tau$ is the total free-evolution time. Assuming a Gaussian noise environment, the coherence function is

$$\xi(\tau) = \tau^2 \int_0^\infty \frac{\mathrm{d}\omega}{2\pi} S_z(\omega) F(\omega,\tau), \tag{46}$$

where $S_z(\omega)$ is the power spectral density of the longitudinal noise that causes the dephasing, and $F(\omega,\tau)$ is a sequence-specific weighting function called the filter function which acts to shape the noise spectrum seen by the qubit [25]. Assuming infinitely short pulses, the filter function for a CPMG sequence with $N$ (even) $\pi$-pulses is $F_{\mathrm{CP}}^{(N)}(\omega,\tau) = 4\operatorname{sinc}^2(\omega\tau/2)\sin^4(\omega\tau/4N)/\cos^2(\omega\tau/2N)$. As illustrated in Supplementary Figure 11, this filter is essentially a bandbpass filter with a passband that peaks around frequency $\omega/2\pi = N/2\tau$, indicating that more $\pi$-pulses will shift the filter to higher frequencies. In addition, the filter bandwidth for a fixed passband becomes narrower with larger $N$.

Taking into account the effect from finite duration of $\pi$-pulses, the modified filter function has a general form [26, 27]

$$F_{\mathrm{CP}}^{(N)}(\omega,\tau) = \frac{1}{(\omega\tau)^2} \left| 1 + (-1)^{1+N}\exp(\mathrm{i}\,\omega\tau) + 2\sum_{j=1}^{N}(-1)^j \exp(\mathrm{i}\,\omega\delta_j\tau)\cos(\omega\tau_\pi/2) \right|^2, \tag{47}$$

where $\delta_j \in [0,1]$ is the normalized position of the center of the $j$th $\pi$-pulse between the two $\pi/2$-pulses and $\tau_\pi$ is the length of each $\pi$-pulse, yielding a total sequence length $\tau + N\tau_\pi$. For $\pi$-pulses of short duration (e.g., $\tau_\pi = 10$ ns) compared with the total free-evolution time (i.e., $N\tau_\pi \ll \tau$), as is typical for our experiment, the bandpass filter frequency still peaks near $\omega/2\pi = N/2\tau$. In practice, we use Eq. (47) to find its precise position.

The photon shot-noise measured in this work has a Lorentzian noise power spectral density centered at zero-frequency. The spectral density is essentially frequency independent at low frequencies (the "white-noise" region of the Lorenzian), and it decreases at higher frequencies (the "tail" of the Lorenzian). For small $N$, such that the filter passband is in the white-noise region, the dephasing time $T_\phi$ does not change with $N$. For large enough $N$, such that the filter passband reaches the Lorentzian tail region, the noise power contributing to the coherence integral in Eq. (46) is reduced. For such large $N$, the dephasing time $T_\phi$ increases as $N$ increases. Since the transverse relaxation time is defined (within a Bloch-Redfield picture) by the rates $1/T_2 = 1/2T_1 + 1/T_\phi$, increasing $N$ in this tail region will extend $T_2$ towards the $2T_1$ limit.

For the qubit described in the main text, $T_{2,\text{CPMG}}$ is approximately $40\,\mu\text{s}$ for $N \leq 100$ and represents a typical duration $\tau$ of the free-evolution. Furthermore, as seen in Fig. 6b in the main text, $T_{2,\text{CPMG}}$ begins to increase around $N = 100$, reaching $T_{2,\text{CPMG}} = 50\,\mu\text{s}$ at $N = 200$. Taking $\tau = 50\,\mu\text{s}$ and $N = 200$, the characteristic frequency is $200/(2 \times 50\,\mu\text{s}) = 2\,\text{MHz}$, consistent with the -3dB point of the Lorentzian spectrum with bandwidth $\kappa/2\pi = 1.5\,\text{MHz}$ (see Fig. 6a in the main text). For $N > 1000$, $T_{2,\text{CPMG}}$ saturates at about $85\,\mu\text{s}$, close to the expected value of $2T_1$.